\documentclass[iop,revtex4]{emulateapj}
\newcommand{\hi}{H\,I}
\newcommand{\HI}{H\,I\,\,}
\newcommand{\NHI}{$\rm{N_{\mathrm{H\,I\,}}}$}
\newcommand{\cmm}{cm$^{-2}\:$}

\newcommand{\kms}{km~s$^{-1}\:$}

\newcommand{\be}{\begin{equation}}
\newcommand{\ee}{\end{equation}}

\newcommand{\VLSR}{\rm{V_{\mathrm{LSR}}}}

\usepackage{graphicx}
\usepackage{amssymb}
\usepackage{amsmath}
\usepackage{hyperref}

\begin{document}

\title{Sensitive 21cm Observations of Neutral Hydrogen in the Local Group near M31}
\shorttitle{Neutral Hydrogen in the Local Group near M31}

\author{Spencer A. Wolfe}
\affil{Dept. of Physics \& Astronomy, West Virginia University, Morgantown, WV 26506}
\email{swolfe4@mix.wvu.edu}
\author{Felix J. Lockman}
\affil{National Radio Astronomy Observatory\altaffilmark{~1}, Green Bank, WV 24944}
\email{jlockman@nrao.edu}
\author{D.J.~Pisano}
\affil{Dept. of Physics \& Astronomy, West Virginia University, Morgantown, WV 26506}
\email{DJPisano@mail.wvu.edu}
\altaffiltext{1}{The National Radio Astronomy Observatory is a facility of the National Science Foundation operated under a cooperative agreement by Associated Universities, Inc.}

\begin{abstract}
Very sensitive 21cm \HI measurements have been made at several locations around the Local Group galaxy M31
 using the Green Bank Telescope (GBT) at an angular resolution of $9\farcm1$, with a  $5\sigma$ detection level 
of \NHI $= 3.9 \times 10^{17}$ \cmm\   for a 30 \kms\  line.  Most of the \HI 
in a 12 square degree area almost equidistant
between M31 and M33 is contained in nine discrete clouds 
that have a typical size of a few kpc and \HI\  mass of $10^5$ M$_{\Sun}$.
Their velocities in the Local Group Standard of Rest  
lie  between $-100$ and +40 \kms,  comparable to the systemic velocities of M31 and M33. The clouds appear to be isolated kinematically and spatially from each other. 
The total \HI mass of all nine clouds is $1.4 \times 10^6$ M$_{\odot}$ for an adopted distance of 800 kpc
with perhaps another $0.2 \times 10^6$ M$_{\odot}$ in smaller clouds or more diffuse emission. 
The \HI mass of each cloud is typically  three orders of magnitude less than the dynamical (virial) mass needed 
to bind the cloud gravitationally.
Although they have the size and \HI mass of dwarf galaxies, the clouds are unlikely to be part of the satellite system 
of the Local Group as they lack stars.  
To the north of M31, sensitive \HI measurements on a 
coarse grid find emission that may be associated with an extension of the M31 high-velocity cloud 
population to projected distances of $\sim 100$ kpc.  An extension of the M31 
high-velocity cloud population at a similar 
distance to the south-east, toward M33, is not observed.  
\end{abstract}


\section{INTRODUCTION}
\label{ch:Intro}

The two largest galaxies in the Local Group, M31 and the Milky Way, have a substantial amount of gas 
residing in a  circumgalactic medium (CGM, also called a gaseous halo), outside of their disks.    Their CGM is 
dominated by ionized gas, but also contains
neutral high-velocity clouds (HVCs) observed in the 21cm line 
\citep{2001ApJS..136..463W, 2003ApJS..146..165S, 2004ApJ...601L..39T, 2008MNRAS.390.1691W, 
2009ApJ...699..754S, Putman2009, 2012MNRAS.424.2896L, Lehner2015}.
Gas likely associated with M31 is seen in absorption against background AGN to projected distances of 
at least 300 kpc   \citep{Lehner2015}.   The Milky Way may 
have a similar CGM less easily separated from disk gas because of projection effects, 
but manifest in the stripping of gas from dwarf spheroidals 
at distances  to  300 kpc \citep{Grcevich2009, 2014ApJ...795L...5S,Gatto2013}.  
 If the CGM of M31 does extend this far, it encompasses
the smaller spiral M33, which itself has a modest population of neutral HVCs \citep{2008A&A...487..161G, Putman2009}.
The CGM of the Milky Way contains the  Magellanic Stream (MS), which extends 
at least $200\arcdeg$ across the sky in \HI\ emission and whose mass 
is probably dominated by ionized gas \citep{2010ApJ...723.1618N, 2014ApJ...787..147F}.  

There are also neutral atomic hydrogen (\HI) clouds in the Local Group whose 
connection with individual galaxies is not understood.  Compact high velocity clouds (CHVCs) and 
ultra-compact high velocity clouds (UCHVCs) are of small angular size and relatively isolated, and 
are candidates for low mass galaxies that may lack star formation entirely 
\citep{2002A&A...392..417D,2013ApJ...768...77A}.   For a 
variety of reasons, CHVCs are now thought to reside in the Milky Way CGM \citep[e.g.,][]{2002ApJS..143..419S}, 
but the location and nature of  the recently-discovered UCHVCs is less clear.

In their study of 21cm emission from \HI in the Local Group, 
\citet[][hereafter BT04]{2004A&A...417..421B}, discovered extended regions of 
 \HI  around the galaxy M31 that formed a partial bridge 
 to the galaxy M33.  This emission was detected at the extremely low levels of 
\NHI $\sim 10^{17}$ \cmm, about two orders of magnitude below the typical 
column density detectable in  
extragalactic 21cm observations \citep{2011A&A...526A.118H}. The BT04 observations
were made with a rather coarse angular resolution of 49\arcmin\  and the origin of the neutral gas was uncertain, 
but the diffuse \HI appeared to connect the systemic heliocentric velocities
of M31 and M33 \citep{Lewis2013}.  
 BT04 proposed that the \HI  arose from condensation in a dark matter-dominated
filament connecting the two galaxies.
Another suggestion was that it resulted from a tidal encounter between the  galaxies \citep{2008MNRAS.390L..24B}.
Subsequent observations of part of the region using the Green Bank Telescope (GBT) at 9\arcmin\  angular 
resolution confirmed the reality of the emission, though those
 data lacked the sensitivity to reveal any detailed structure
\citep{2012AJ....144...52L}.  As the existence and properties of CGM and 
intra-group gas is critical to our understanding of the
 formation and evolution of galaxies \cite[e.g.][]{2001ApJ...552..473D,Fukugita2006,ChenHW2010,Putman2012,Cen2013,Lehner2015}, 
we have undertaken a major survey 
of the area around and between 
M31 and M33 using the Green Bank Telescope, which provides both the sensitivity needed to detect this 
extremely faint emission, and the angular resolution to discern some of  its structure.

In a previous paper \citep[][hereafter Paper I]{2013Natur.497..224W} we presented results of the first part of the study, 
which showed that a significant fraction of the \hi\ detected in a 12 square-degree field 
south-east of M31 in the direction of 
M33 arose in discrete structures which, assuming they are 800 kpc distant, have the size of dwarf galaxies but 
 no detectable stellar 
component.  Subsequently, \citet{2013ApJ...776...80M} suggested that 
there may be a stellar overdensity in the direction of 
one of the clouds, but the association between the gas and stars is not established at this time.
The clouds have velocities  
similar to the systemic velocities of M31 and M33 and are thus probably not part of the high-velocity 
cloud (HVC) system of either galaxy.  Here we report on additional observations
 that reveal more clearly  the structure and content of the \hi\ clouds, as well as new observations of selected 
directions to the north of M31.  The distances to M31 and M33 are
$\sim$ 750 and $\sim$ 850 kpc, respectively \citep[e.g.][]{2012ApJ...745..156R, 2010ApJ...715..277B} 
so we assume that the material located between them lies at a distance of 800 kpc.  

\section{OBSERVATIONS}

All data used here were obtained with the 100-meter Robert C. Byrd Green Bank Telescope \citep[GBT,][]{2009IEEEP..97.1382P}
 with the dual-polarization 
L-band receiver that has a total system temperature of $\lesssim18$ K at elevations
$>20\arcdeg$.  At the frequency of the 21cm line, the telescope has a half-power beam-width of 
$9\farcm1$.  Spectra were measured using the GBT Spectrometer, which provides a total velocity coverage 
$>1000$ \kms\ in the 21cm line at a channel spacing of 0.32 \kms.  Spectra were calibrated using observations 
of 3C48 and the antenna response analysis from \cite{2011A&A...536A..81B}. Velocities were measured with respect to the 
Local Standard of Rest (LSR), but 
in this part of the sky, $\VLSR$  differs from heliocentric velocities by only a few \kms.

\begin{figure}
\includegraphics[width=\columnwidth]{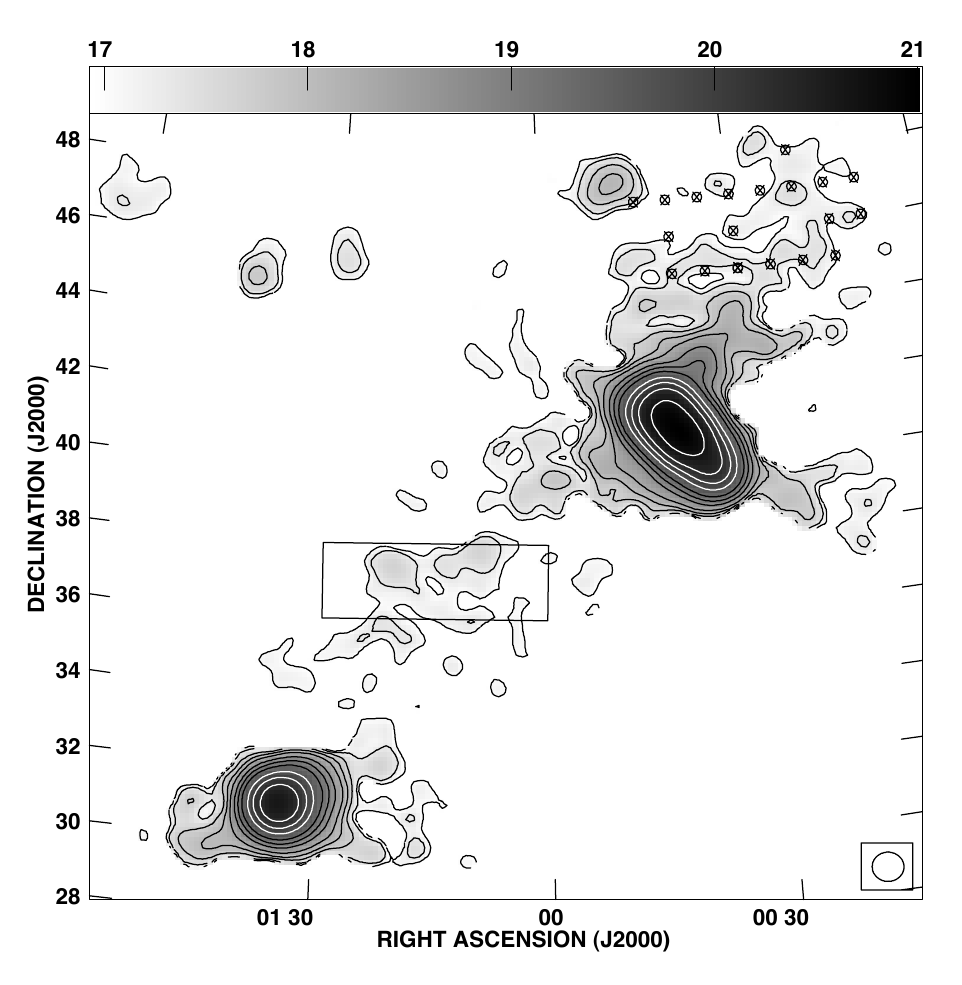}
\caption{Locations of the current observations marked on a map of the total \NHI\ from  
the BT04 survey.  The galaxies M33 and M31 lie at the lower left and upper right, respectively.
Contours are for $\rm{log(N_{HI}) =}$ 17.0, 17.3, 17.7, 18.0, 18.3, 18.7, 19.0, 19.3, 19.7, 20.0, 20.3 and 20.7 $\rm{(cm^{-2})}$. 
Our  ``M31-M33'' field, outlined with a box,  lies between 
the two galaxies and was covered with complete sampling.  North of M31, we made 
a series of pointed observations at locations marked with boxed crosses. The GBT angular resolution
is $9\arcmin$, but the boxed crosses are $15\arcmin$~in size to make them easier to see.
The boxed circle in the lower right shows the resolution of BT04.} 
\label{fig:ObservedPositions}
\end{figure}

Figure~\ref{fig:ObservedPositions} shows the observed areas on the BT04 \HI map of the region.  
Data reduction was done with special procedures written in GBTIDL \citep{2006ASPC..351..512M}, the stray radiation  
correction and calibration followed \citet{2011A&A...536A..81B}, and special procedures were written to remove instrumental 
baselines from spectra in the data cube.  

\subsection{The M31-M33 Field}

As described in Paper I, we used the GBT to map a $6\arcdeg \times 2\arcdeg$ field in right ascension and declination 
centered at $\rm{J2000 = 01^h16^m +37\arcdeg00\arcmin}$.  The observations were made while moving the telescope in right 
ascension at a fixed declination, binning the data in 3-second
 samples every 1\farcm6 in right ascension.  At the end of the 6\arcdeg\ strip the telescope was stepped in
declination by 3\farcm6 and the scanning direction was reversed. This procedure 
covered the area with Nyquist sampling in 
declination, and finer sampling in the moving coordinate to prevent beam broadening \citep{2007A&A...474..679M}.  
Regions at eastern edge of each row  -- areas identified as having no emission by BT04 and 
confirmed from our observations -- were used as reference positions.  Spectra from the eastern-most 16\arcmin\  of
each strip  were averaged and 
supplied the reference spectrum for the rest of the spectra in the strip.  
As the reference spectrum had an integration time of 30 seconds, this 
greatly reduced the noise in the difference spectrum. 
The field was observed over and over again until the desired noise level was reached.  In all, the M31-M33 field was 
observed for about 
400 hours  with an average time per GBT beam of 46 min. Our observing procedure could cancel some emission, but 
only if it had a nearly constant amplitude and $\VLSR$ over scales of many degrees.  We see no evidence of this in the data.
There are three positions in our map where \citet{2012AJ....144...52L} made frequency-switched detections of \HI emission associated with the clouds. 
Those data are in reasonable agreement with our current measurements.  At another position where only a small upper limit was reported, we likewise 
see no emission. This indicates that our position-switched technique has not cancelled significant amounts of \HI. In addition, the current data are consistent 
with the sensitive GBT spectrum reported in BT04, but note that the declination reported in BT04 has a typographical error: the correct position of the GBT 
spectrum is $\rm{J2000 = 01^h20^m29\arcsec +37\arcdeg22\arcmin33\arcsec}$.

Spectra were corrected for atmospheric attenuation and a second-order polynomial was fit to emission-free 
channels to provide statistics for a quick check on data quality.  In general, instrumental baselines
were exellent and modeled well by a 2nd or 3rd order polynomial.  
A small fraction of the spectra ($4\%$) was rejected for having poor instrumental baselines 
caused mainly by radio frequency interference or temporary instrumental effects.
 The spectra were smothed to an effective velocity resolution of 5.15 \kms\ and 
gridded into a cube using \emph{AIPS}\footnotemark \footnotetext{Developed by the National Radio Astronomy Observatory: 
\href{http://www.aips.nrao.edu/index.shtml}{http://www.aips.nrao.edu/index.shtml.}} with a pixel spacing of $1\farcm75$ 
using a spherical Bessel interpolation function following \citet{2007A&A...474..679M}.  
A third-order baseline was removed from each spectrum in the cube.   The procedure of subtracting 
a nearby reference position should effectively remove stray radiation
\citep{2011A&A...536A..81B}  hence no further corrections were applied. 
The noise in the final cube varies slightly with position, with a typical 
value, in   brightness temperature, of 
 $\rm{\sigma_T} = 3.45$ mK in a 5.15 \kms\ channel. 
This gives a $5\sigma$ limit on \NHI\ of $3.9 \times 10^{17}$ for a 30 \kms\ FWHM line. The equivalent \ion{H}{1} mass limit is $\rm{\sim 10^4~ M_\Sun}$, 
assuming a distance of 800 kpc. We note that previous surveys around galaxy groups typically have mass limits $\rm{\gtrsim 10^5~M_\Sun}$ 
\citep{2006MNRAS.371.1617A} or $\rm{10^6~M_\Sun}$ \cite[e.g.][]{2001MNRAS.325.1142Z, 2007ApJ...662..959P, 2008AJ....135.1983C}.

\subsection{M31 North Observations}

Observations were also made at 18 positions north of M31 to investigate the nature of the very faint 21cm emission
detected by BT04 in this area.  The observed positions are shown 
in Figure~\ref{fig:ObservedPositions}.   Here, because reference positions were not readily available, 
the data were taken by  frequency switching between the 21cm rest frequency and a band 
 4 MHz (844 \kms) away within the 12.5 MHz band of the GBT Spectrometer. 
Data were calibrated and corrected for stray radiation as described by \citet{2011A&A...536A..81B}, 
 and a third or fourth order polynomial was fit to emission-free  velocities. 
The data were then smoothed from 0.3 \kms velocity resolution to 1.3 \kms.
The typical on-source integration time for each pointing is 76 minutes and the 
median noise in a 1.3 \kms\ channel is 3.7 mK.   This gives a $5\sigma$ limit on 
\NHI\ of $1.6 \times 10^{17}$ \cmm\ for a spectral line with a full width at half maximum (FWHM) of 30 \kms.

\section{NEUTRAL HYDROGEN BETWEEN M31 AND M33}

\begin{figure*}
\includegraphics[width=\textwidth]{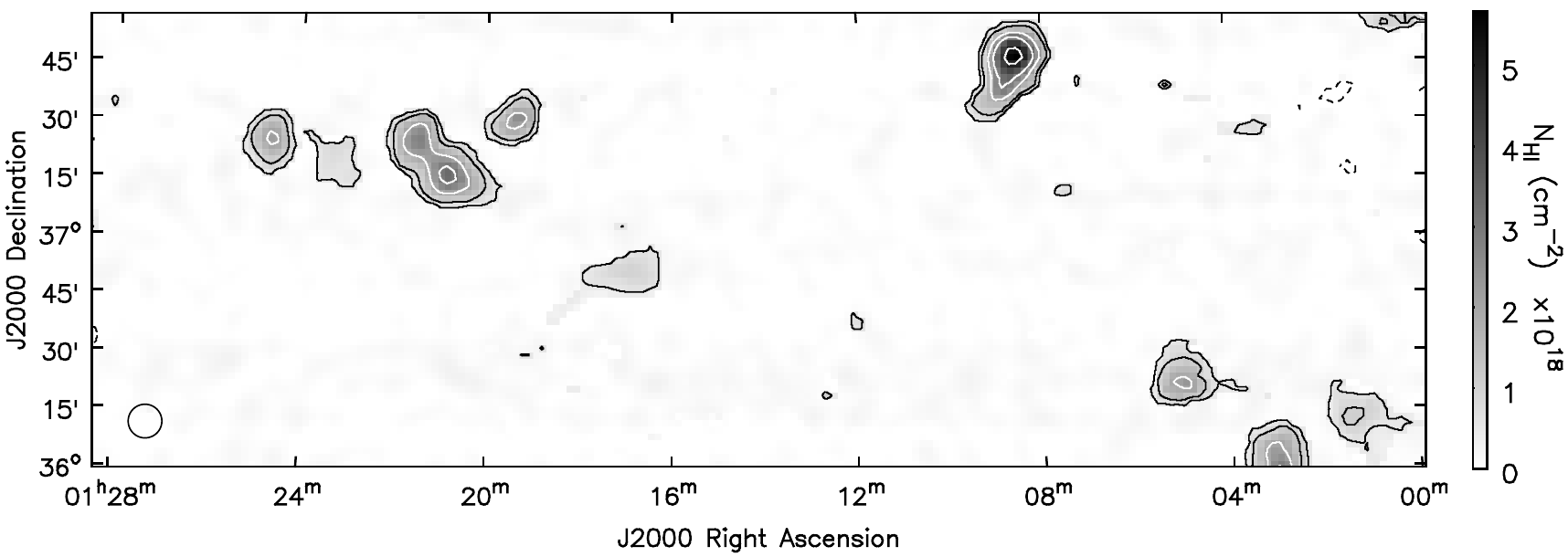}
 \caption{Integrated \HI\ column density map of the  M31-M33  clouds over $\rm{-359 \leq V_{LSR} \leq -187}$ \kms.
The contours are at -1, 1, 2, 4, 6 and 10 times increments of 5$\times 10^{17}$ \cmm. 
The circle in the lower left shows the angular resolution of the GBT.}
\label{fig:M0-map} 
\end{figure*}

\begin{figure}
\includegraphics[width=\columnwidth]{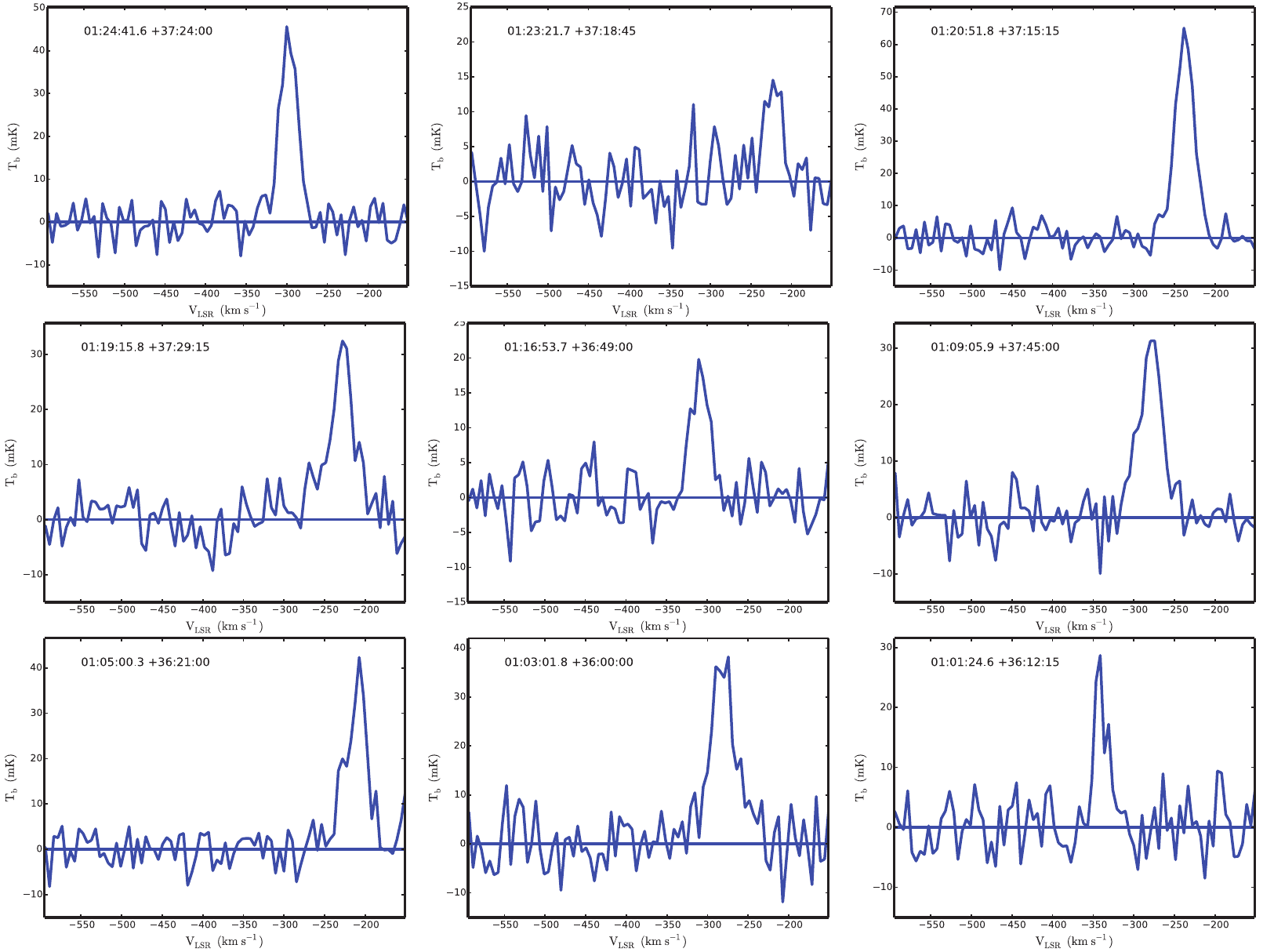}
 \caption{Spectra taken at the peak $\rm{N_{HI}}$ of the nine clouds detected in the M31-M33 field. The vertical axis is 21cm 
brightness temperature and the horizontal axis is $\VLSR$.}
\label{fig:spectra}
\end{figure}

An integrated intensity map of the spectra summed over $-359 \leq \VLSR  \leq -187$ \kms\ and
converted to column density, \NHI, is presented in Figure~\ref{fig:M0-map}. 
This range encompasses all detected \HI emission not associated with the Milky Way. 
Emission from the disks of M31 and  M33 overlaps that from the Milky Way at some $\VLSR  \geq -150$ 
\kms, but these velocities are not in the range studied here.
\NHI~is calculated under the assumption that the 
emission is optically thin, an excellent assumption for lines with T$\rm{_b} < 100$ mK.    
The rms noise in column density is $1.9 \times 10^{17}$ \cmm\ ($1\sigma$);  
contours are drawn in multiples of $5 \times 10^{17}$ \cmm.
The \HI\ emission in this field is dominated by discrete clouds, some of which are resolved by the GBT 
9\farcm1 beam.  Six of these clouds were detected at full angular resolution in Paper I, the other three 
appear in the more sensitive data presented here.
Spectra toward the peak \NHI\  of each cloud are shown in Figure~\ref{fig:spectra}.  

A Gaussian function was fit to the 21cm spectrum of each cloud at the location of its peak \NHI, and 
the components, along with the location of the peak, are given in Table~\ref{tab:properties}.  
Errors are $1\sigma$ from the Gaussian fit.  Clouds are numbered for convenience; the identification is 
usually not the same as in Paper~I.

\begin{deluxetable*}{cccccc}
\centering
\tablewidth{\textwidth}
\tablecaption{Clouds Detected Between M31 and M33 
\label{tab:properties}}
\tablehead{
\colhead{Cloud} &\colhead{J2000}  & \colhead{${\rm T_{\rm L}}$}  & \colhead{FWHM} & \colhead{$\VLSR$} &
\colhead{${\rm N_{HI}}$} \\
\colhead{} & \colhead{(hh:mm:ss dd:mm:ss)}  & \colhead{(mK)} & \colhead{(\kms)} & 
\colhead{(\kms)}   &
\colhead{($10^{18}~\rm{cm^{-2}}$)} \\
\colhead{(1)} & \colhead{(2)} & \colhead{(3)} & \colhead{(4)} & \colhead{(5)} & \colhead{(6)}  \\
}
\startdata
1 & 01:24:41.6  +37:24:00  & 44.0$\pm2.4$  & 25.5$\pm$1.6 &$-297.8\pm0.7$  & $2.2\pm 0.1$ \\
2 &  01:23:21.7 +37:18:45  & 10.9$\pm2.2$  & 39.3$\pm9.3$ & $-223.0\pm4.0$ & $0.8\pm 0.1$ \\
3 & 01:20:51.8  +37:15:15  & 63.2$\pm1.8$  & 27.2$\pm0.9$ & $-237.3\pm0.4$ &$3.3\pm0.1$ \\
4 & 01:19:15.8  +37:29:15  & 28.5$\pm2.1$  & 38.1$\pm$3.3 & $-228.2\pm1.4$  & $2.2\pm 0.1$ \\
5 & 01:16:53.7  +36:49:00  & 18.3$\pm1.7$  & 26.2$\pm$3.0 & $-308.7\pm1.2$ & $0.9\pm0.1$ \\
6 & 01:08:29.6  +37:45:00   & 81.3$\pm3.6$ & 32.0$\pm$1.9 & $-278.6\pm0.6$ & $5.0\pm 0.1$\\
7 & 01:05:00.3  +36:21:00   & 34.7$\pm3.0$ & 31.4$\pm$3.5 & $-210.2\pm1.2$  &$2.1\pm0.2$ \\
8 & 01:03:01.8  +36:00:00   & 35.0$\pm2.6$ & 36.8$\pm$3.4 & $-281.4\pm1.3$ &$2.5\pm0.2$ \\
9 & 01:01:24.6  +36:12:15  &  25.4$\pm3.2$ & 19.2$\pm2.9$ & $-341.0\pm1.2$ & $0.9\pm0.1$ \\ 
\enddata
\end{deluxetable*}

\begin{deluxetable*}{ccccccccc}
\centering
\tablewidth{\textwidth}
\tablecaption{Derived Cloud Properties\tablenotemark{a}
\label{tab:derived_properties}}
\tablehead{
\colhead{Cloud} &\colhead{${\rm V_{LGSR}}$} & \colhead{Diam\tablenotemark{b}}  & \colhead{${\rm \langle Diam \rangle}$\tablenotemark{c}} & 
\colhead{${\rm \langle FWHM \rangle}$\tablenotemark{d}}  & \colhead{$\rm{r_{1/2}}$\tablenotemark{e}} 
 & \colhead{${\rm M_{HI}}$} &  \colhead{$\rm{M_{dyn}}$\tablenotemark{f}} & \colhead{$\rho$\tablenotemark{g}} \\
\colhead{} &  \colhead{(\kms)} & \colhead{(kpc)} & \colhead{(kpc)}  &  \colhead{(\kms)} & \colhead{(kpc)} & 
\colhead{(${\rm 10^{4}\  M_{\odot}}$)} & \colhead{ (${\rm 10^{8}\  M_{\odot}}$)} & \colhead{(kpc)}    \\
\colhead{(1)} & \colhead{(2)} & \colhead{(3)} & \colhead{(4)} & \colhead{(5)}  & \colhead{(6)}  & \colhead{(7)} & \colhead{(8)} & \colhead{(9)}  \\
}
\startdata
1 & $-61$      & 4.4         & 3.2  & $22.2\pm0.6$ &  0.38: &  $12.7\pm0 .2$ &  0.5: & 126 \\
2 &  +14          & 4.4         & 2.8  & $28.0\pm 3.5$ &   0.75 &  $4.5\pm 0.2$ & 1.5 & 123 \\
3 &  +1            & 7.2        & 5.3   & $27.6\pm 0.8$  &  &    $33.0\pm 0.3$  & & 118 \\
3a &               &                  &           &  $28.0\pm1.0$  &  0.82 &  $22.7\pm0.2$ & 1.6 & \\
3b &            &                 &            &  $27.1\pm1.3$ &   0.34:   &   $10.2\pm0.2$  &  0.6: & \\
4 &  +12          & 3.7        & 3.1   & $26.5\pm 1.5$      &   0.41: &  $8.6\pm 0.2$ & 0.7: & 112\\
5 &  $-69$    &  7.0       & 3.8  & $21.8\pm2.5$      &     1.13 &    $7.8\pm  0.2$ &   1.3 & 109\\
6 &  $-32$     & 7.2        & 4.6  &   $33.6\pm 1.3$  &    0.78 &  $39.2\pm 0.3$  &  2.2 & 87 \\
7 &  +36           &  4.9       & 3.9    &  $27.4\pm 2.1 $ &   0.71 &  $12.6\pm 0.4$  & 1.3 & 92\\
8 &  $-35$ &  $>3.5$ & $>3.2$             & $29.6\pm 2.1$  &  &  $>11.6$  &   & 91\\
9 & $-94$      &  3.5        & 3.2 & $20.5\pm2.3$       & 0.96  &    $8.7\pm 0.2$    & 1.0 & 88\\
\enddata
\tablenotetext{a}{For an assumed distance of 800 kpc.}
\tablenotetext{b}{Maximum cloud extent.}
\tablenotetext{c}{Diameter of a circle with an area equal to that of the cloud.}
\tablenotetext{d}{Column density weighted average FWHM over the entire cloud.}
\tablenotetext{e}{Square root of the product
of the major and minor axis radii.}
\tablenotetext{f}{From Equation~\ref{eq:dynmass}.}
\tablenotetext{g}{Projected distance from M31.}
\end{deluxetable*}

Table~\ref{tab:derived_properties} gives derived properties of the clouds.
  The velocity with respect to the Local Group (V$\rm{_{LGSR}}$) was determined
 using the calculator given in NED\footnote{The NASA/IPAC Extragalactic Database 
is operated by the Jet Propulsion Laboratory, California Institute of Technology, under contract with the National Aeronautics and Space Administration.}.  
The quantity {\it Diam}  is the maximum cloud extent  measured down to the $3\sigma$ noise  
level while $\langle Diam \rangle $ is the diameter
of a circle with an area equal to that of the cloud.  The difference between these quantities is 
a measure of the elongation of a cloud.   The ``average FWHM'' of the 21cm line averaged over the entire 
cloud is given in Col.~5.  Cloud 8 was not completely mapped so some of its quantities are limits.
There is an additional emission feature at the very north-west edge of the field, 
J2000 $= 01^h00^m24^s +37\arcdeg53\arcmin$ 
with $\rm{T_L} \approx 10$ mK at $\VLSR = -264$ \kms\ and a 
peak \NHI $\lesssim 10^{18}$ \cmm, but we cannot characterize it further.

The  \HI mass in Col.~7 was calculated by integrating over 
 velocities relevant to each cloud, then summing  
over an area around the cloud. The mass 
assumes that the emission is optically thin.  Errors on the mass are derived 
from the noise in emission-free channels over the area  of each cloud.

A two-dimensional Gaussian was fit to each cloud yielding a major and minor axis size.  For this, 
the two parts of Cloud 3 were treated separately.  Major and minor axis radii from the Gaussian fitting 
were then deconvolved to produce an estimate of the true angular size given the 9\farcm1 beam of the GBT, which is approximately Gaussian: 
$\rm{r_{true} = {(r_{obs}^2 - 4\farcm55^2)}^{1/2}}$.  
The minor axis of Clouds 1, 3b and 4 were unresolved within the errors, and for these we assume an 
intrinsic radius of 1\arcmin, equivalent to 233 pc 
at the assumed 800 kpc distance of the clouds. Quantities derived from this adopted radius are 
consequently uncertain, and are marked with a colon (:) in Table~\ref{tab:derived_properties}.
The square root of the product of the deconvolved major and minor axis radii is given
in Col.~6 of Table~\ref{tab:derived_properties} and called  $\rm{r_{1/2}}$.
This is the average radius within which half the 
\HI mass is contained as estimated from the deconvolved Gaussian fit. 

 Using this radius
we calculate a dynamical (or virial)  mass -- the total mass (from whatever source) needed to bind a
cloud of radius  $\rm{r_{1/2}}$ that has a given velocity dispersion.  
We note that the definition of the {\it dynamical} or {\it virial} or {\it total} mass derived 
from the size and velocity structure of an object differs considerably from author to author, 
and an exact determination requires  
information that we do not possess, such as  the density structure within a cloud.  
For simplicity, and to allow comparisons with other measurements (see \S \ref{ch:discussion}) we 
adopt the following \citep{2008gady.book.....B}:
\begin{equation}
   \rm{M_{dyn} \equiv  \frac{2~ r_{1/2}~ \sigma_v^{2}}{G}},
\end{equation}
 where G is the gravitational constant and $\rm{\sigma_v}$ is the 3-dimensional velocity dispersion of the cloud.
As $\rm{\sigma_v}$ is not a measurable 
quantity, we use the FWHM from the measured average spectrum, and assume isotropy to calculate 
\begin{equation} 
   \rm{  \left(  \frac{M_{dyn}}{M_\odot}  \right)  \equiv  2.5 \times 10^5 \left( \frac{r_{1/2}}{kpc}  \right)   \left(\frac{FWHM}{km~s^{-1}} \right)^2,   }
\label{eq:dynmass}
\end{equation} 
 where $\rm{r_{1/2}}$ is in kpc, FWHM in \kms and M$\rm{_{dyn}}$ is in solar masses.  
Dynamical masses are given in Col. 8.   In all cases 
they exceed the observed \HI mass by a factor of $\sim 10^3$. Col. 9 gives $\rho$, the projected distance
  from M31  assuming that the clouds are at a distance of 800 kpc. 
The clouds lie 
$\gtrsim$ 90 kpc away from M31's center, farther than the distances of the known HVCs around M31, 
all of which have $\rho < 50$ kpc \citep{2008MNRAS.390.1691W}.

\subsection{The Velocity Range}

In the field of Figure~\ref{fig:M0-map}, we find no \HI\ emission at velocities more negative than that associated 
with cloud 9, i.e., nothing at $\VLSR \lesssim -370$ \kms and no emission at the more positive velocities 
between $-190 \lesssim \VLSR \lesssim -160 $ \kms.  
At still more positive $\VLSR$, there is a band of \HI\ extending from
 the south center of the field to the north-west that is quite bright by our standards  (T$\rm{_b} > 0.6$ K) 
at $-150 \lesssim  \VLSR \lesssim -100 $ \kms.  This appears to connect smoothly 
to Milky Way emission at more positive velocity and we will not consider it further here.

\citet{Lehner2015} argue that the CGM of M31 should be defined 
as having $-300 \lesssim \VLSR \lesssim -150 $ \kms, 
that velocities more negative than this may arise in the Magellanic Stream and velocities more
positive in the halo of the Milky Way.  Our Cloud 9 has a peak \NHI\ at $\VLSR = -341$ \kms, but is in no way 
unusual in its size, mass, linewidth, or projected distance from M31, so we assume 
that it is part of the cloud population.   
\citet{2008MNRAS.390.1691W} have detected 21cm emission that they attribute 
to M31 HVCs only $\sim 2\arcdeg$ to the northwest of the  Figure~\ref{fig:M0-map} field 
at velocities as negative as $\VLSR \approx -500$ \kms. 
 We would have easily detected   
similar emission in the Figure~\ref{fig:M0-map} 
field.   Apparently the M31 HVC population does not extend over this area.  
This will be discussed further in \S \ref{ch:discussion}.

\subsection{The Total Neutral Hydrogen Mass}
\label{sec:neutralmass}

Figure~\ref{fig:M0-map} shows some evidence for \HI\ emission outside of the clouds listed in 
Table~\ref{tab:properties}, but this is relatively small and ususally concentrated in discrete regions. 
Assuming that all the \HI we detect is at a distance of 800 kpc,
the sum of the \HI\  mass of all nine clouds is  $1.4 \times$ 10$^6$ M$\rm{_{\odot}}$, while integration over the entire
field yields $1.6 \times$ 10$^6$ M$_{\odot}$.  Thus  virtually all of the 
neutral gas we measure is contained in the nine clouds with about half  in just 
the two largest clouds, numbers 3 and 6. 

\begin{figure}
\includegraphics[width = \columnwidth]{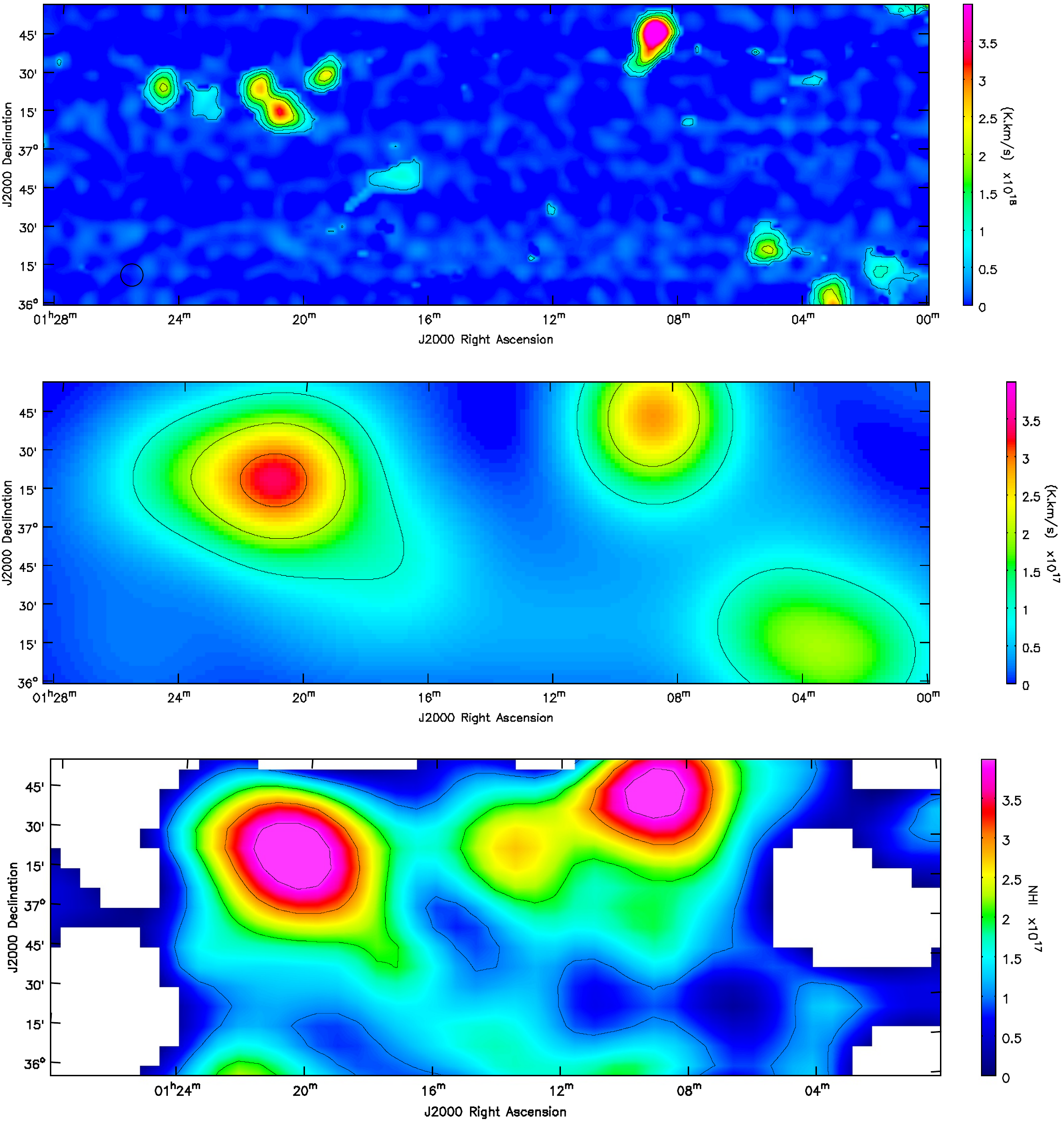}
 \caption{Top panel: integral column density of the M31-M33 field at full GBT angular resolution; contours 
are every $5 \times 10^{17}$ \cmm\ begining at $5
\times 10^{17}$ cm$^{-2}$ and the color scale runs from zero to $4 \times 10^{18}$ \cmm.
Middle panel: GBT data smoothed to the $49\arcmin$ angular resolution of the BT04 
measurements using the WSRT beam from \citet{2008A&A...479..903P}.  For this and the lower panel, 
the color scale runs from zero to $4 \times 10^{17}$ \cmm, and the contours are every $10^{17}$ cm$^{-2}$ beginning at $10^{17}$
cm$^{-2}$.
Lower panel: BT04 integral column density map at $49\arcmin$ resolution.
Regions without detectable \HI are blanked.
While many of the general features of our GBT 
data agree with those from BT04, the BT04 map contains regions of emission not found in the GBT data.
 }
\label{fig:convolved_m0}
\end{figure}

To compare our data with those of BT04, we match the angular resolution of that survey by convolving 
the integrated spectra  shown in Figure~\ref{fig:M0-map} to an angular resolution of $49\arcmin$ using the WSRT beam from \citet{2008A&A...479..903P}. 
 Results are shown in Figure~\ref{fig:convolved_m0}.  
While the maps have many similarities there are significant differences.  The most striking is near 
J2000 = $01^h13^m30^s, +37\arcdeg24\arcmin$, where the BT04 data show emission 
that does not appear in the GBT data.  
The BT04 survey also contains emission at the south-central part of our field, some of which may arise from a cloud 
 just off our map at $01^h10^m, +35\arcdeg30\arcmin$ convolved with the much larger BT04 beam. 
In some areas we find good agreement between the two surveys -- 
the northwest and northeast corners, for example --
but overall BT04 reports  an \HI mass $\rm{2.5 \times 10^6 ~M_\odot}$ 
for the area, while we find only $63\%$ of this amount.

We believe that the cause of this discrepancy lies mostly, though not entirely, in the choice of velocity range 
of integration for the BT04 survey. 
No emission is detected with the GBT at velocities more negative than $-370$ \kms.
 If the GBT data are integrated over all velocities $\leq -150$ \kms, the total 
mass remains $1.6 \times$ 10$^6$ M$_{\odot}$, but if the upper limit is taken to be $ -140$ \kms, 
only 10 \kms\  more positive, the 
total \HI mass doubles.   We discuss above the reasons for not including emission at $\VLSR \geq -150$ \kms\ 
in the census of Local Group gas, as it likely arises in the Milky Way (\citet{Lehner2015} 
conclude this as well).  It is 
plausible that some of this emission is present in the BT04 map, 
especially as those spectra had a rather coarse velocity 
resolution of 17 \kms.   This explanation does not account for the 
discrepancy with the BT04 feature at $01^h13^m30^s +37\arcdeg24\arcmin$.
The GBT spectrum in this direction is entirely consistent with noise at $\VLSR  \leq  -145$ \kms.

 In Paper I we reported that there was $\approx1 \times 10^6$  M$_{\odot}$ of \HI\ 
in the GBT data that was not associated with discrete clouds.  We now believe that this 
conclusion is incorrect and 
resulted from a very small systematic baseline error with an amplitude of only a few mK in the preliminary 
GBT data.  In the current data there can be no more than $0.2 \times 10^6$ M$_{\sun}$ of \HI outside the 
nine clouds.  We do not believe that our data reduction procedure is artifically suppressing real emission; 
instrumental baseline fitting removes only low-order polynomials preserving lines of normal velocity width, 
and the position-switching observing 
technique preserves structure on angular scales $\lesssim 5\arcdeg$.
If there were emission in the reference regions beyond the edge of the map, we would see it as negative
 features in the map.  As no such features exist, we have
 confidence in our estimates of M$_{\rm HI}$.

\subsection{Velocity and Linewidth}

\begin{figure}
\includegraphics[width = \columnwidth]{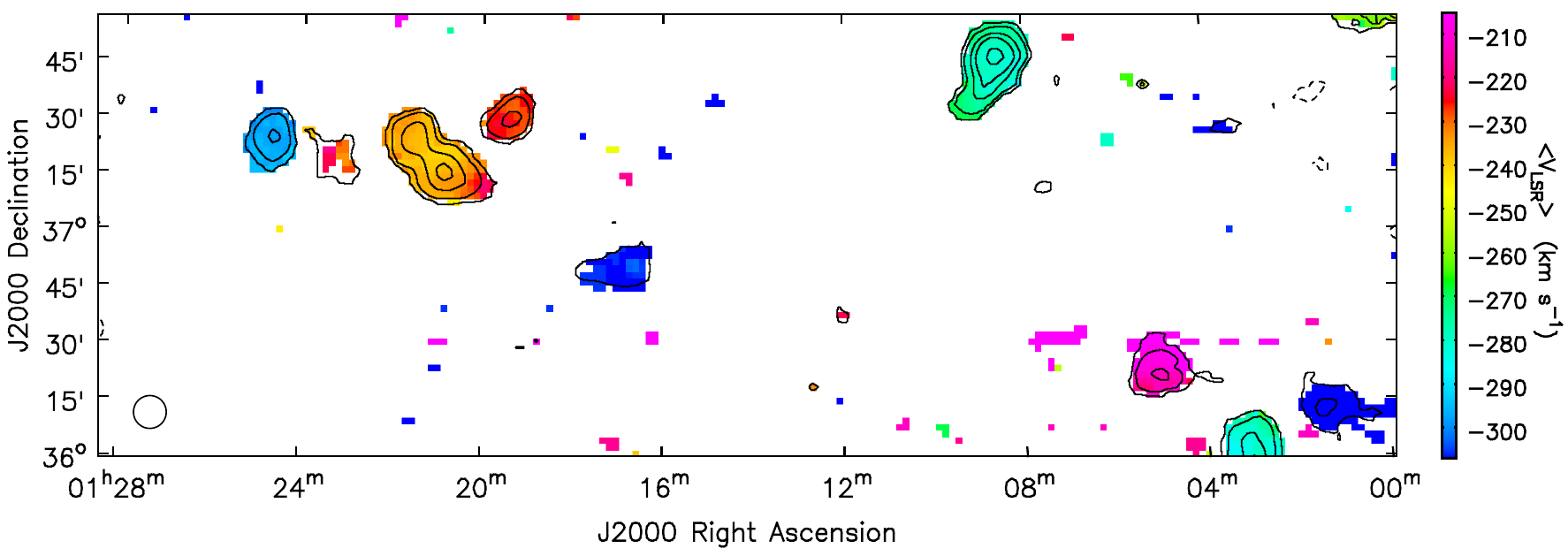}
 \caption{Average V$_{\rm LSR}$ over the field, calculated for channels between $\rm{-359 \leq V_{LSR} \leq -187}$ 
\kms\  that have $\rm{T_b > 4\sigma =  0.015}$ K.   
Although velocities can vary significantly from cloud to cloud,  the mean 
velocity within a cloud is relatively constant. The circle in the lower left shows
the angular resolution of the GBT. It is apparent that the clouds are separated not only spatially but kinematically.}
\label{fig:M1-map}
\end{figure}

\begin{figure}
\includegraphics[width = \columnwidth]{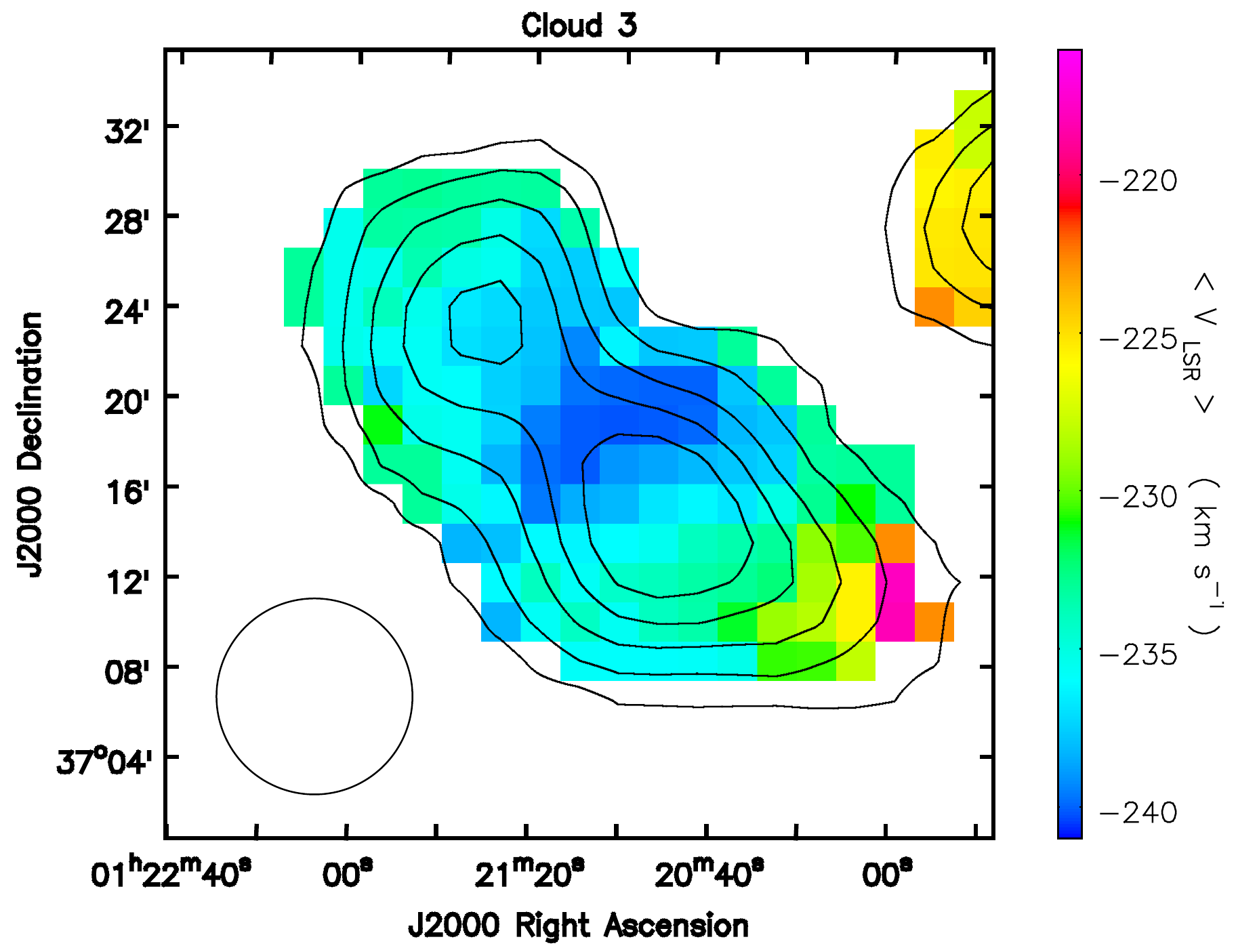}
 \caption{Average V$_{\rm LSR}$ pixel-by-pixel for Cloud 3 together with contours of \NHI.  
Averages were calculated only for channels with $\rm{T_b > 4\sigma = 15}$ mK.
Contours are at 1, 2, 3, 4, and 5 times increments of $5 \times 10^{17}$ \cmm.
Unrelated emission from Cloud 4 appears in the upper right. The circle in the lower left shows
the angular resolution of the GBT. Both cloud components show a 10 \kms velocity gradient between their center and edge.
 }
\label{fig:Cloud-2_M1}
\end{figure}

\begin{figure}
\includegraphics[width = \columnwidth]{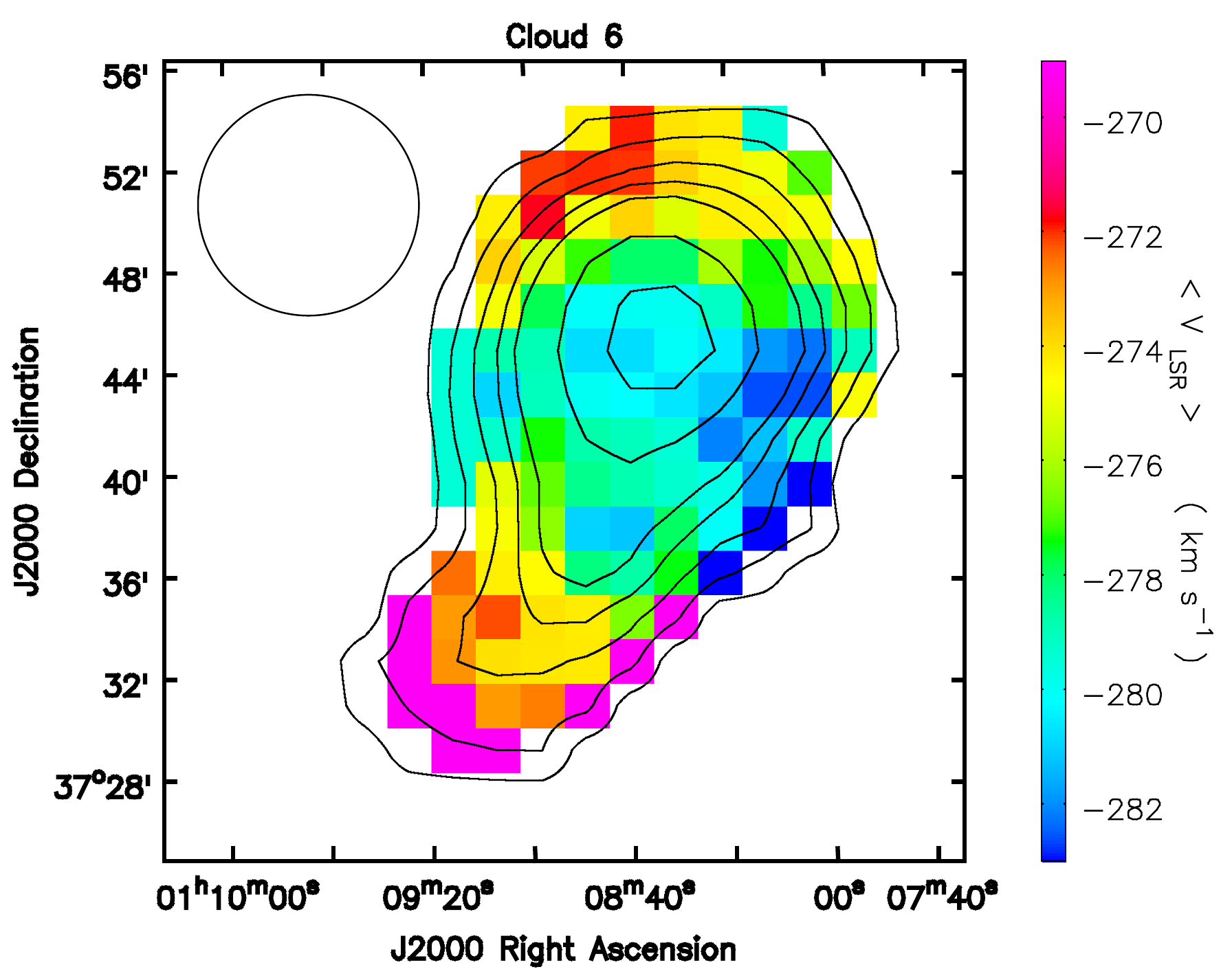}
 \caption{Average V$_{\rm LSR}$ pixel-by-pixel for Cloud 6 together with contours of \NHI.  
Averages were calculated only for channels with $\rm{T_b > 4\sigma = 15}$ mK.
Contours are at 1, 2, 3, 4, 5, 7.5 and 10 times increments of $5 \times 10^{17}$ \cmm.  
The circle in the upper left shows
the angular resolution of the GBT.  This cloud shows a 10 \kms velocity gradient from the 
center to its edges.
}
\label{fig:Cloud-5_M1}
\end{figure}

The average velocity of the \HI\ emission  is shown pixel-by-pixel 
 in Figure~\ref{fig:M1-map}.  Contours are the same as in
 Figure~\ref{fig:M0-map}.
Here, to give an indication of the overall velocity pattern of the clouds the 
velocity-weighted values of the brightness temperature were
calculated over the entire range of detectable emission: $-359$ to $-187$ \kms, clipping the data
at the approximate $4\sigma$ noise level of 15 mK.  
The clouds span a range of velocity between $-341 \leq \VLSR \leq -210$ \kms.  In our 
data, there are only two clouds that have significant internal velocity structure.  These are shown in 
Figures~\ref{fig:Cloud-2_M1} and~\ref{fig:Cloud-5_M1}.

Cloud 3 consists of two components with nearly identical velocities at their peak, but overall there is a 10 \kms\ 
gradient from $-240$ \kms\ between the two cloud components to $-233$ \kms\ at the northern edge and 
$-230$ \kms\ at the southern edge.  Cloud 6 has a head-tail appearance with, once again, velocities most negative 
at the center and rising by $\approx 10$ \kms\ at the northern and southern edges.  

The mean linewidth at the peak N$_{\rm HI}$ of the nine clouds is $27.9\pm4.3$ \kms, essentially identical to the 
median.  The lines are well-fit by single Gaussians.   Unlike Galactic HVCs and CHVCS, there is no evidence 
for two components in any of the lines as would be expected if they have a two-phase temperature structure
\citep{2002ApJS..143..419S}.   
The FWHM does not vary much across an individual cloud.  The 
FWHM averaged over the entire cloud (Table~\ref{tab:derived_properties}) 
and toward the peak \NHI\  (Table~\ref{tab:properties}) are quite similar.  The range of linewidths is 
19.2 - 32.5 \kms, much smaller than the range observed in 17 M31 HVCs (at essentially identical angular 
resolution) of 11 - 71 \kms \citep{2008MNRAS.390.1691W}. 

Figure~\ref{fig:M1-map} re-enforces the impression from 
Figure~\ref{fig:M0-map}  that the individual clouds are independent entities,  as 
nearby clouds can have quite different velocities.  Clouds 7 and 9, for example,  
differ by 120 \kms yet are separated by  $<1\arcdeg$ on the sky ($\sim 10$ kpc at 800 kpc distance).

\section{NEUTRAL HYDROGEN TO THE NORTH OF M31}

\begin{figure}
\includegraphics[width = \columnwidth]{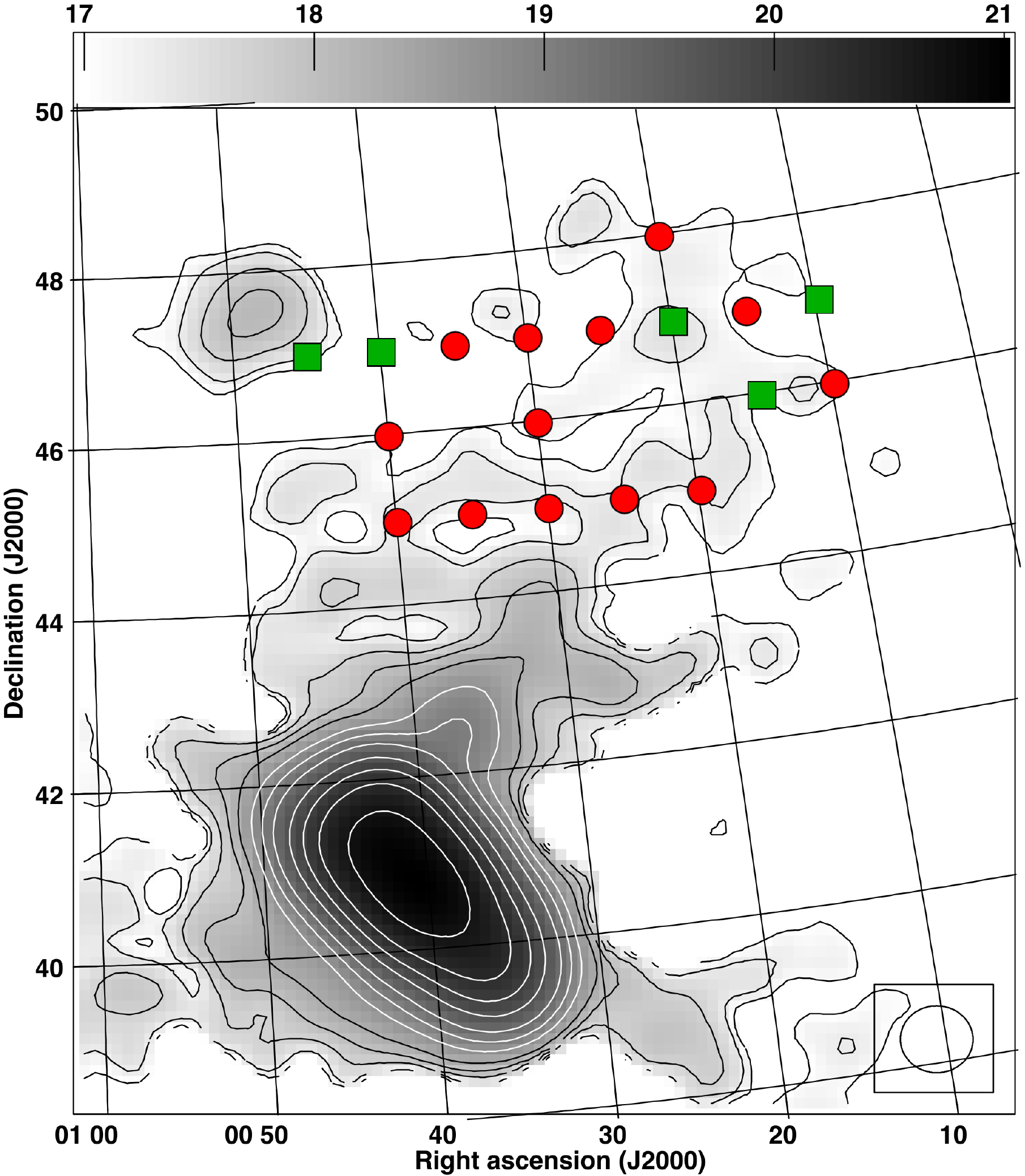}
\caption{Expanded view of the BT04 survey \HI  data from Figure~\ref{fig:ObservedPositions} to show the locations of the
 GBT measurements to the north-west of M31. The GBT angular resolution is 9\arcmin, but the symbols are 15\arcmin~across 
for easier identification. 
Green boxes mark positions with detected \HI emission
at $\VLSR \leq -150$ \kms. Red circles show positions without a detection.
The circle in the lower right shows the BT04 angular resolution.
}
\label{fig:POINTINGS}
\end{figure}

Results from the measurements to the north of M31 are given in Table~\ref{tab:northclouds}, 
and illustrated in Figure~\ref{fig:POINTINGS} as symbols on a map of the BT04 survey.  
We include only emission with $\VLSR \leq -150$ \kms. 
Line properties were derived from a Gaussian fit, and 
errors are $1\sigma$.  Values of \NHI\ come also from the Gaussian fit; 
here errors reflect both the noise and an assumed equal contribution from baseline uncertainties.  
The median $5\sigma$ sensitivity to a 30 \kms ~FWHM line is $1.6 \times 10^{17}$ \cmm.
Because we have only incomplete sampling, it is not possible to delineate objects and 
calculate a mass or size.  The values of \NHI\ should be understood as random samples of the 
medium and not peak values, in contrast to the values given in Table~\ref{tab:properties} for  the M31-M33 field.

\begin{deluxetable*}{cccccccccc}
\tablewidth{\textwidth}
\tablecaption{Observations North of M31
\label{tab:northclouds}}
\tablehead{
 \colhead{J2000}  & \colhead{Galactic} & \colhead{$\rm{{\sigma_T}}$} &  \colhead{T$_{\rm L}$} &\colhead{$\VLSR$} & 
\colhead{FWHM} &  \colhead{$\rm{N_{HI}}$} & \colhead{$\rm{V_{LGSR}}$} &\colhead{$\rho$} \\ % & \colhead{Notes} \\
 \colhead{(hh:mm dd:mm)} &   \colhead{l$^{\circ}$, b$^{\circ}$} & \colhead{(mK)}  & \colhead{(mK)} & \colhead{ (\kms)} & \colhead{ (\kms)} & \colhead{$({ \rm 10^{17}~cm^{-2}})$} &   \colhead{(\kms)} & \colhead{(kpc)} \\
\colhead{(1)} & \colhead{(2)} & \colhead{(3)} & \colhead{(4)} & \colhead{(5)} & \colhead{(6)} & \colhead{(7)} & \colhead{(8)} & \colhead{(9)}
}
\startdata
 00:10  +46:00  & & 3.8  &  &  & & & &\\
 00:10  +47:00  & 115.6, -15.3&4.4 & $27.5\pm1.3$  & $-176\pm1$  & 25.9 $\pm$ 1.4  & $13.8\pm 1.0$ & 106 & 115  \\
 00:15  +46:00  & 116.3, -16.4&3.5 &$15.6\pm1.2$  & $-238\pm1$ & 16.0 $\pm$ 1.4 & $4.8\pm 0.6$  & 43 & 96 \\
                          & &3.5 & $22.4\pm1.2$ & $-388\pm1$ & $23.6\pm1.4$ & $10.2\pm0.8$ & -109 &\\
 00:15  +47:00  & &3.3   &   &   & && & \\
 00:20  +45:00  &  &6.4   & & & \\
 00:20  +47:00  & 117.4, -15.5 &3.3 & $30.4\pm1.0$ & $-166\pm1$ & 26.1 $\pm$ 1.0 & $15.4\pm0.7$ & 112 & 98 \\
 00:20  +48:00  & &3.6   &   &   & & & &\\
 00:25  +45:00  &  &3.8   &  &  & & & & \\
 00:25  +47:00  & &3.7    &   &   &  & & & \\
 00:30  +45:00  & &4.1   &   & & & & &\\
 00:30  +46:00  & &4.4         &  &  &  & & &\\
00:30  +47:00  & &4.5  &     &   & &  & & \\
00:35  +45:00  & &4.1  &  &   &  & & &\\
 00:35  +47:00  & &3.7  &  & & & & &\\
 00:40  +45:00  & &3.3 &  & & & & & \\
 00:40  +46:00   & &3.8 &  & & & & & \\
 00:40  +47:00  & 120.9, -15.8 &2.9  & $18.3\pm0.7$ & $-182\pm1$  & $25.1\pm1.1$ & $8.9\pm0.6$ & 85  & 81 \\
 00:45  +47:00  & 121.8, -15.9 &3.7  & $18.8\pm0.8$ & $ -183\pm1$  & $42.5 \pm$ 2.1  & $15.5 \pm1.0$ & 86 &  80 \\
\enddata
\end{deluxetable*}

Our results are rather puzzling in view of the BT04 data.  We detect no emission at
$\delta = +45\arcdeg$, 
at only one position at $\delta = +46\arcdeg$,  and yet have four detections at $\delta = +47\arcdeg$, 
two of which lie outside the lowest BT04 contour.   
The first conclusion must be that the \HI\  is much patchier than would be inferred from BT04, which 
does not give a good representation of the \HI\ at $\VLSR \leq -150$ \kms.  
In this sense the M31 north measurements complement the conclusion from the M31-M33 
field (Figure~\ref{fig:convolved_m0}).
In the GBT data, bright 
lines  identified with Milky Way emission are regularly found 
 at $\VLSR \approx -100$ \kms.  It is possible that the BT04 map includes some of this material, 
which might account for the discrepancy with the GBT measurements.  
Our detection of \HI outside the BT04 contours at $\delta = +47\arcdeg$ would 
imply that we are seeing very small angular sized features that suffer from
 beam dilution in the BT04 measurements.  This was also a conclusion from Paper~I.

\begin{figure}
\includegraphics[width = \columnwidth]{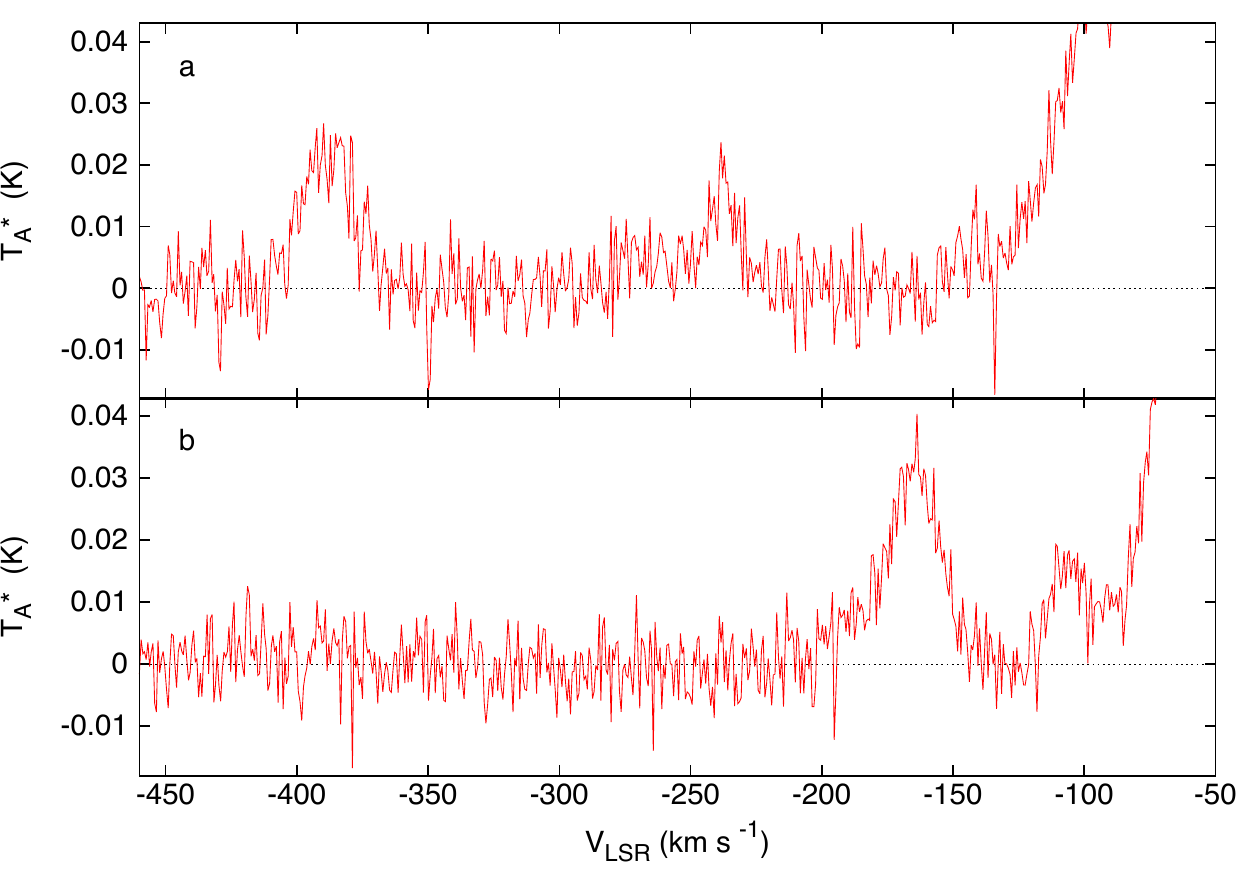}
 \caption{[a] Spectrum from a deep GBT pointing toward J2000 = $00^h15^m +46\arcdeg00\arcmin$. 
The component at $-238$ \kms\  has a velocity similar to M31's HVCs while the component
 at $-388$ \kms\ is closer to the systemic velocity of M31. [b] Spectrum from a pointing toward J2000 = $00^h20^m +47\arcdeg00\arcmin$. 
The spectral line lies $\sim$ 50 \kms~ from Milky Way emission.
 }
\label{fig:POINTSPEC}
\end{figure}

The position at $00^h15^m, +46\arcdeg00\arcmin$ has two 21cm line components 
separated by 150 \kms. Figure~\ref{fig:POINTSPEC}a shows this spectrum 
as well as the spectrum at  $00^h20^m, +47\arcdeg00\arcmin$
  (\ref{fig:POINTSPEC}b).  Both illustrate 
 the emission near $-100$ \kms\ that we attribute to a component of the Milky Way.

\section{KINEMATICS OF THE EMISSION}
\label{ch:kinematics}

\begin{figure}
\includegraphics[width = \columnwidth]{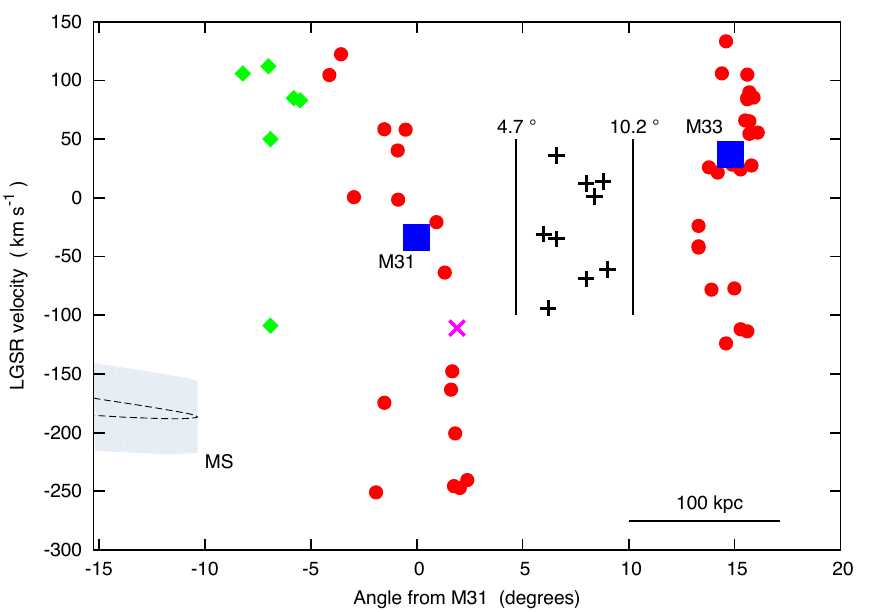}
 \caption{Velocity with respect to the 
Local Group Standard of Rest ($\rm{V_{LGSR}}$) versus angular distance from M31, 
where directions toward M33 are taken to be positive.
 The blue squares are M31 and M33, with the red dots being the high velocity cloud populations of
   each galaxy. The black crosses are the clouds from the M31-M33 map with  
 vertical lines marking the map limits. The green diamonds are detections from the GBT pointings 
north of M31. The pink cross marks the UV absorption line measurements of \citet{2015arXiv150600635K} towards 
a background AGN.
The MS curve is the estimated position of Stream S0 from \citet{2010ApJ...723.1618N}, 
with $\pm$ 30 \kms spread to account 
for the velocity width of the Stream and our approximations. 
In most cases the Stream lies hundreds of \kms away from the emission we detect. The rotation of 
M31 extends from -353 to +248 \kms in $\rm{V_{LGSR}}$, and for M33 from -105 to +158 \kms 
\citep{Braun2009,Putman2009}.
 }
\label{fig:LGSR}
\end{figure}

Velocities of all GBT detections are shown in Figure~\ref{fig:LGSR} as the Local 
Group Standard of Rest velocity (V$_{\rm LGSR}$) versus angular distance from M31. 
We include both the clouds of  Figure~\ref{fig:M0-map} and the detections to the north of M31. 
For comparison, the known HVCs around M31 and M33 are indicated with red circles 
\citep{2008MNRAS.390.1691W,2008A&A...487..161G, Putman2009}
and the systemic velocities of M31 and M33 with blue rectangles \citep{1996AJ....111..794K}. 
 As discussed in Paper I, the clouds between M31 and M33 lie at velocities similar to the systemic velocity of 
both galaxies, and at a larger distance from either galaxy than their HVCs. This is true also of the two 
newly-identified clouds between M31 and M33. 
Most of the emission to the north of M31 appears consistent with arising in 
an extension of the M31 HVC population. 
It is in the velocity range associated by \citet{Lehner2015} with the circumgalactic medium of M31.  

\citet{2015arXiv150600635K}  report detection of gas possibly associated 
the the M31 Stellar Stream through UV absorption line 
measurements against background AGN.   Their most secure
detection is in lines of Si, C, and O against Q0045+3926 
 at a location about $2\arcdeg$ from M31 in the direction of M33.  They find one system at $\VLSR = -370$ \kms, or 
${\rm V_{LGSR} = -111}$ \kms, which is marked with a pink cross in Figure~\ref{fig:LGSR}.  
It lies in the region occupied by M31 HVCs.  

The second \HI\ component toward  $00^h15^m, +46\arcdeg00\arcmin$ (Figure~\ref{fig:POINTSPEC}a) 
 at $\VLSR = -388$ \kms\ (V$_{\rm LGSR} = -109$ \kms)
 has a velocity that might be associated with 
the Magellanic Stream \citep{Lehner2015}
though this direction lies nearly $20\arcdeg$ from the axis of the 
Magellanic Stream as defined by \citet{2010ApJ...723.1618N}, and 
not far from a direction that does not show UV absorption arising in the Stream 
\citep{2014ApJ...787..147F}.  
As our measurements are only over a sparse grid in this area, 
we do not know the size or mass of any cloud associated with this feature.
One of the M31 HVCs, the Davies Cloud, has also been 
suspected of being part of the Stream because its mass would be an order of magnitude larger than the other 
M31 HVCs were it at the distance of M31 \citep{1975MNRAS.170P..45D, 2008MNRAS.390.1691W}. The Davies Cloud 
does not have extraordinary kinematics and does not occupy an unusual location in Figure~\ref{fig:LGSR}.  
At the distance of M31 its \HI mass is similar to that of the Milky Way HVC Complex C \citep{Wakker2007}. 
 We see no reason to exclude it from the M31 HVC population.

Because \citet{2010ApJ...723.1618N} have shown that the Magellanic Stream (MS) extends into the 
BT04 area, 
we have tried to estimate where the MS would lie in Figure~\ref{fig:LGSR}. 
Five \HI Streams were identified in the BT04 data from \citet[][Figure 4]{2010ApJ...723.1618N}, labeled S0-S4.
 We take Stream S0, the one closest to our pointings, to lie along a line of constant RA at $23^h45^m$ and choose a declination range from +20 to +50 degrees. 

To approximate the LGSR velocities of Steam S0 along this line, we first transform 
the positions to Magellanic Longitude \citep[$\rm{L_{MS}}$,][]{2008ApJ...679..432N} and use the extended
$\rm{V_{LSR}}$ vs. $\rm{L_{MS}}$ fiducial curve from Figure 7b of \citet{2010ApJ...723.1618N},
 converting to V$_{\rm LGSR}$. 
The resulting angle-velocity for Stream S0 is shown as a dashed curve in  Figure~\ref{fig:LGSR}. 
To account for the spread in velocity of the Stream, and for our approximations, we over-plot 
a shaded region with bounds of $\pm$ 30 \kms from the curve. 

The point of closest approach of S0 to M31 is at an angular distance of $10.3\arcdeg$  
with $\rm{V_{LGSR}}$ = -187 \kms. 
Most of our detections to the northwest of M31 lie at least 225 \kms from the 
average velocity of S0 ($\rm{V_{LGSR}} \approx$ -175 \kms). 
The line at $00^h15^m +46\arcdeg00\arcmin$, V$_{\rm LGSR} = -109$ \kms\ is the only one within $5\arcdeg$ and 50 \kms of S0, 
so excepting this, it seems unlikely that the emission we detect arises in the Magellanic Stream.

\section{DISCUSSION: THE ORIGIN OF THE CLOUDS BETWEEN M31 AND M33}
\label{ch:discussion}

In the 12 square-degree area between M31 and M33 observed with the GBT,  
 the faint \HI\ discovered by BT04 is resolved into at least nine discrete clouds.   
Here we consider possibilities for the origin of the clouds.

\subsection{Products of Galaxy Interactions}

Although BT04 surveyed a large area, they  found \HI\ at relevant velocities only near  
a line joining {M31 and M33} on the sky (Figure~\ref{fig:ObservedPositions}),  
and not widespread in the Local Group.   
This naturally leads to models 
where the \HI\ tracks past  tidal interactions, as in the M81 group \citep{1994Natur.372..530Y, 2008AJ....135.1983C}.  
\citet{2008MNRAS.390L..24B} modeled the BT04 results as arising from a close encounter between M31 and M33 
some 4-8 Gyr ago, and others have also considered an 
interaction a few Gyr ago as an explanation for features in  M33's 
extended HI distribution and the stellar structures in M31 and M33 
 \citep{Putman2009,2009Natur.461...66M,Lewis2013}. 
The Magellanic Stream is a good, nearby template of an ongoing galaxy interaction 
\citep[e.g.,][]{Besla2010,Stanimirovic2010} but its age is quite uncertain, in the range of 0.3 - 2.0 Gyr \citep{2015arXiv151105853D}.
The \HI\ mass and average \NHI\ is generally higher than we observe between M31 and M33 
\citep{Stanimirovic2008,2008ApJ...679..432N,2010ApJ...723.1618N}.  
It is worth noting that as gaseous tidal debris ages it should 
disperse causing its \NHI\ to decrease, so the M31-M33 clouds may represent material that is 
substantially older than is seen in the Magellanic Stream.
 
A recent study of the evolution of the Local Group has now cast doubt 
on the possiblity that we are observing the relic of a past interaction between M31 and M33,
as it indicates that in the last $\approx12$ Gyr, M33 has not been closer to M31 than it is now
\citep{2013MNRAS.436.2096S}.  
The same calculations though,  suggest that the dwarf galaxies And II and And XV could have 
interacted with M31 over the last 0.7 Gyr creating an \HI extension toward M33.  
Since Local Group dwarfs with detectable \HI\
typically have M$\rm{_{HI}}$/L$\rm{_V \sim}$1 \citep{2012AJ....144....4M, 2014ApJ...795L...5S}, we can estimate how much \HI\ And II and And XV
could have contributed to the M31 CGM.  Based on the M$\rm{_V}$ from \citet{2012AJ....144....4M}, their total \HI\ mass would be $\rm{\sim 10^7 ~M_\Sun}$, which 
is comparable to the total \HI\ mass in the entire filamentary structure between M31 and M33.  
It seems unlikely, however, that the encounter would leave the nine clouds we detect, each apparently coherent, but spread over 
a projected distance of 70 kpc with a spread in $\rm{V_{LGSR}}$ of 130 \kms.
It is clear that more detailed modelling of the Local Group, including its gas, is needed to address 
these possibilities.

\subsection{Dwarf Galaxies}

The larger M31-M33 clouds have an \HI mass similar to that of some dwarf galaxies, but 
the clouds are not likely to be associated with stellar systems.  
Table~\ref{tab:dwarfs} compares properties of the most massive cloud with 
dwarf galaxies of a similar \HI mass.  
The dwarf spheroidal And XII is included as it is one of the faintest of the known M31 satellites, 
but was detected easily in a recent search for even fainter systems \citep{2013ApJ...776...80M}. 
To allow for accurate comparisons we calculate dynamical masses from the measured quantities in Table~\ref{tab:dwarfs}, using Eq.~\ref{eq:dynmass}.

\begin{deluxetable*}{cccccccc}

\tablewidth{\textwidth}
\tablecaption{Comparison of an M31-M33 Cloud with Dwarf Galaxies
\label{tab:dwarfs}}
\tablehead{
\colhead{Object} & \colhead{r$_{1/2}$} & \colhead{FWHM\tablenotemark{b}} & \colhead{${\rm M_{HI}}$} &  \colhead{${\rm M_{*}}$} & 
\colhead{${\rm M_{V}}$}  & \colhead{$\rm{M_{dyn}}$} & \colhead{References}\\
\colhead{} &  \colhead{(kpc)} & \colhead{(\kms)} & \colhead{(${\rm M_{\odot}}$)} & \colhead{(${\rm  M_{\odot}}$)} & 
\colhead{(mag)}  & \colhead{(${\rm M_{\odot}}$)} \\
\colhead{(1)} & \colhead{(2)} & \colhead{(3)} & \colhead{(4)} & \colhead{(5)}  & \colhead{(6)} &
 \colhead{(7)} & \colhead{(8)}\\
}
\startdata
Cloud 6\tablenotemark{a} & 0.78 & 34 &  $3.9 \times 10^5$  & ---                        & --- &  $2.2 \times 10^8$ & a \\
Leo P     & 0.25 & 24 & $9.5 \times 10^5$   & $5.7 \times 10^5$ & $-9.4$ & $3.6 \times 10^7$ & b,c,d \\
Leo T    &  0.17 & 16 & $2.8 \times 10^5$  & $1.4 \times 10^5$ & $-8.0$  & $1.1 \times 10^7$ & e,f,g,h\\
And XII &  0.30    & 6  &      ---                  & $3.1 \times 10^4$ & $-6.4$ & $2.8 \times 10^6$ & i,j\\
\enddata
\tablenotetext{a}{For an assumed distance of 800 kpc.}
\tablenotetext{b}{Of the HI emission except for And XII where it is from the stars.}
\tablerefs{ (a) This Work; (b) \citet{Bernstein-Cooper2014}; (c) \citet{2013ApJ...768...77A}; (d)  
\citet{2013AJ....146..145M}; (e) \citet{2008MNRAS.384..535R}; 
(f) \citet{2007ApJ...670..313S}; (g) \citet{2013ApJ...777..119F}; 
 (h) \citet{2008ApJ...680.1112D}; (i) \citet{2010MNRAS.407.2411C}; (j) \citet{2012AJ....144....4M}}
\end{deluxetable*}

Local Group dwarfs with detectable \HI typically have a stellar mass similar to their gas mass 
\citep{2014ApJ...795L...5S}. 
For Cloud  6 this would imply $\rm{M_V} < -8$, and its stars would certainly  have 
been detected already.  \citet{2013ApJ...776...80M} report a possible stellar feature at the Cloud 6 
position, but is considerably fainter than And XII, and if it is actually associated with the Cloud, would 
imply M$_{\rm HI}$/M$_* > 10$.  The dynamical mass of Cloud 6 is also about an order of magnitude higher than 
that of Leo P, even though Leo P has $\sim10$ times more baryonic mass. 
Using the relationship between total mass and M$\rm{_V}$ from the Local Group data of 
\citet{2012AJ....144....4M}, Cloud 6 -- indeed most of the clouds detected in \HI between M31 and M33 -- 
should have a stellar counterpart with M$\rm{_V} < -13$, whereas the 
surveys of stars near M31 suggest M$\rm{_V} > -6$.  
 If Cloud 6 is  a galaxy it is one with rather extreme properties.  It will be interesting to see if 
the possible association with a slight stellar overdensity discovered by \citet{2013ApJ...776...80M}  
reveals a real stellar component of this cloud.
The dwarf galaxy Leo P has a small rotational velocity, $15\pm5$ \kms, comparable to its velocity dispersion 
\citep{Bernstein-Cooper2014}.  As shown in Figure~\ref{fig:Cloud-5_M1} ,  Cloud~6 shows a slight gradient in $\VLSR\ $
from center to edge; at our angular resolution any rotational component is comparable to or smaller than its velocity dispersion.

Another reason why the M31-M33 clouds are not likely to be galaxies is that the stellar 
satellites of M31 and the Milky Way that lie closer than $\approx300$ kpc to 
the parent galaxy are extremely deficient in \HI, 
with mass limits typically well below $10^4$ M$_{\sun}$.  In some cases M$_{\rm HI} \leq 100$ M$_{\sun}$ 
\cite[][R. Beaton private communication]{Grcevich2009, 2014ApJ...795L...5S}.    
The two M31 satellite galaxies that lie nearest the M31-M33 field, And II with $\VLSR = -187$ \kms, 
and And XV with $\VLSR = -322$ \kms, are among those that lack detectable \HI  emission \citep{2012AJ....144...52L}.
Apparently M31 and the Milky Way are very efficient at stripping gas from small satellites passing through 
their CGM \citep{Mayer2006,Grcevich2009,2011MNRAS.415..257N,Gatto2013}.

\begin{figure}
\includegraphics[width = \columnwidth]{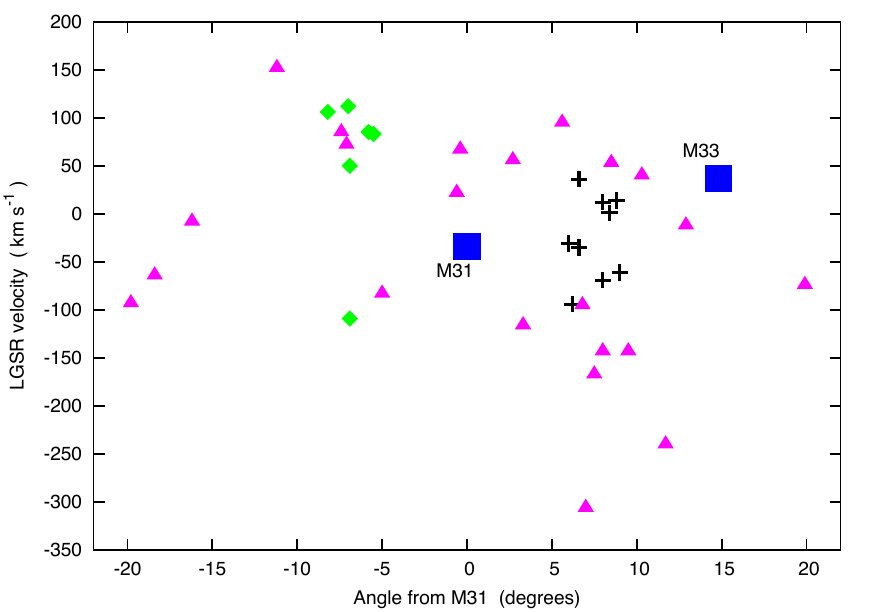}
 \caption{Velocity with respect to the 
Local Group Standard of Rest ($\rm{V_{LGSR}}$) versus angular distance from M31, 
where directions toward M33 are taken to be positive.
 The blue squares are M31 and M33, and the black crosses are the clouds from the M31-M33 map. The green diamonds are detections from the GBT pointings 
north of M31.  The pink triangles are dwarf galaxies from the compilation of 
\citet{2012AJ....144....4M}.  There is no apparent connection between the dwarf galaxies and 
the M31-M33 \HI clouds.
 }
\label{fig:dwarfs}
\end{figure}

Figure~\ref{fig:dwarfs} contains the same \HI data as Figure~\ref{fig:LGSR} but instead of the M31 and M33 HVCs, 
the location and velocity of dwarf galaxies from \citet{2012AJ....144....4M} are shown.  There is no obvious 
connection between the dwarf galaxies and the \HI clouds as the dwarfs are spread over 400 \kms while the clouds have an average 
V$\rm{_{LGSR}}$ like that of M31 and M33 and a total range of only 130 \kms.

\subsection{High Velocity Clouds and Dark Matter Sub-Halos}

Table~\ref{tab:HVCs} gives information that allows us to compare the M31-M33 clouds with various 
populations of HVCs. For the M31-M33 clouds, values are the median from Table~\ref{tab:derived_properties}. 
For the M31 HVCs the listed radius is the median found by \citet{2005A&A...432..937W} 
while the FWHM and ${\rm M_{\HI}}$  are the median values from \citet{2008MNRAS.390.1691W}.  
Values for the Ultra-Compact HVCs (UCHVC) are from \citet{2013ApJ...768...77A} scaled to 
a distance of 800 kpc.  

\begin{deluxetable*}{cccccc}
\tablewidth{\textwidth} 
\tablecaption{Comparison M31-M33 Clouds with HVCs
\label{tab:HVCs}}
\tablehead{
\colhead{Object} & \colhead{r$_{1/2}$} & \colhead{FWHM} & \colhead{${\rm M_{HI}}$} 
 & \colhead{$\rm{M_{dyn}}$} & \colhead{References}\\
\colhead{} &  \colhead{(kpc)} & \colhead{(\kms)} & \colhead{(${\rm M_{\odot}}$)} & 
 \colhead{(${\rm M_{\odot}}$)} \\
\colhead{(1)} & \colhead{(2)} & \colhead{(3)} & \colhead{(4)} & \colhead{(5)}  & \colhead{(6)} \\
}
\startdata
M31-M33 Clouds & 0.75 & 27 & $1.2 \times 10^5$ & $1.4 \times 10^8$  & a \\
M31 HVC    & 0.52 & 26 & $4.7 \times 10^5$    & $8.8 \times 10^7$ & b, c \\
UCHVC   &  1.16 & 23 & $1.2 \times 10^5$    & $1.5 \times 10^8$ & d \\
\enddata
\tablerefs{ (a) This Work; (b) \citet{2005A&A...432..937W}; (c) \citet{2008MNRAS.390.1691W}; (d) \citet{2013ApJ...768...77A} }
\end{deluxetable*}

There is considerable overlap between the physical properties of the M31-M33 clouds 
 and the M31 HVCs, but much less so between their locations and kinematics.
Many HVCs are located close to the edge of the M31 disk  in intersecting filaments \citep{2005A&A...432..937W} .  
Although median line widths are similar, the M31 HVCs have widths as low as half and as high as 
twice the FWHM of the M31-M33 clouds; $90\%$ have a FWHM in the range 13-67 \kms, whereas the 
entire sample of M31-M33 clouds has FWHMs between 19 and 39 \kms.
The  HVCs also have a median mass four times larger than the clouds.   Indeed, there is no cloud in 
our sample that has an \HI\  mass as high 
as the median \HI\ mass of the M31 HVCs.    
Most importantly, the kinematics of the clouds differ significantly from 
that of the  HVCs near them (Figure~\ref{fig:LGSR}), and  they have a substantially smaller spread in velocity 
(see also Paper I). With only one exception the detections north of M31 have positions and velocities consistent with being part of the M31 HVC population
while the nine clouds between M31 and M33 do not.  
The M31 HVCs in the direction of M33 have considerably more negative velocities than the systemic velocity of 
M31. For M31  ${\rm V_{LSGR} = -34}$ \kms, while its HVCs towards M33 have ${\rm \langle V_{LGSR}\rangle = -166}$ \kms.
In contrast, the M31-M33 clouds have ${\rm \langle V_{LGSR}\rangle = -25}$ \kms.    The clouds do not appear to be  simply an extension 
of the M31 HVC population.
Still, 
the physical properties displayed in Table~\ref{tab:HVCs} suggest that the two populations may be formed 
by similar processes, and that the UCHVCs, if at a similar distance, have similar properties as well.

\citet{2004ApJ...601L..39T} and \citet{2005A&A...432..937W} 
considered whether the HVCs of M31 might be the baryonic component of 
 a population of low-mass dark matter halos, an idea with a long history in the study of HVCs  
\citep{1966BAN....18..421O, 1999ApJ...514..818B, 1999A&A...341..437B, 2002A&A...392..417D,
2014MNRAS.442.2883N}.   Of particular importance is that some HVCs (and 
some dwarf galaxies) show a two-component structure in their 21cm lines, suggesting the 
presence of  gas in equilibrium at two temperatures.  This is a diagnostic of pressure, and was used (among other 
pieces of evidence) to argue that the compact HVCs (CHVCs) must lie 
at distances $\lesssim 150$ kpc from the Milky Way, and not at distances of $\sim1$ Mpc 
\citep{2002ApJS..143..419S}.  The failure to detect CHVC
analogs spread throughout other groups of galaxies 
also implies that they must be located within 90 kpc of individual galaxies \citep{2007ApJ...662..959P}.  This result is 
consistent with HVC detections around other galaxies such as NGC~891 and NGC~2403 \citep{2007AJ....134.1019O, 2002AJ....123.3124F} 
and the distance brackets for Milky Way HVCs \cite[e.g.][]{2001ApJS..136..463W, 2007ApJ...670L.113W, 2008ApJ...672..298W}.

The UCHVCs are discussed in depth in \citet{2013ApJ...768...77A}  and \citet{2013ApJ...777..119F}.  
The dwarf galaxy Leo P was originally classified as a UCHVC on the basis of its 21cm characteristics 
before its stellar component was detected and it was discovered to be at a distance 
of 1.7 Mpc \citep{2013AJ....146...15G,2013AJ....145..149R,2013AJ....146..145M}.
The \HI  properties listed in Table~\ref{tab:HVCs} for the UCHVCs show that they are similar to 
the M31 HVCs and the M31-M33 clouds.   There is no evidence for a two-phase 
interstellar medium in the M31 HVCs, the M31-M33 clouds, or in the unidentified UCHVCs, 
although it is found in other HVCs and the Magellanic Stream \citep{Kalberla2006,Stanimirovic2010}.
This
is important as it implies that these objects exist in regions of low external pressure \citep{2002ApJS..143..419S}. 
The observations to date, however, have relatively poor linear resolution and might not be able to 
detect a cool \HI phase if present in moderate amounts.  

\subsection{Gas in Planes of Satellites or a Dark Matter Filament}

Lacking an estimate of the distance to the clouds, it is not clear if they are 
related to the planes of satellite galaxies that are now thought to 
be fundamental structures in the Local Group \citep{2013Natur.493...62I, 2012ApJ...758...11C, 2013ApJ...766..120C}.
The M31-M33 field studied here lies in the part of the sky where M31 satellites are 
aligned in the structure called ``Plane 2''.   It lies to the west of M31 and 
extends southwards towards M33 \citep{2013MNRAS.436.2096S}.  
The two M31 satellite galaxies nearest to our field on the sky have velocities within the 
range of the \HI\ clouds (see Figure 3 of \citet{2013MNRAS.436.2096S})
 and it is an intriguing possibility that the M31-M33 clouds are part of a larger alignment of matter in this part 
of the Local Group.  Although study of these planes is only beginning, if they represent large-scale 
dark matter structures, then it is very plausible that they would be accompanied by enhancements 
in gas density.

BT04 suggested 
that the \HI\ they discovered  originated from condensation of hot 
gas in a dark matter filament connecting M31 and M33. This is
now testable, in part, through cosmological simulations that attempt to 
follow the evolution of systems like the Local Group. 
\citet{2014MNRAS.441.2593N} and \citet{2015A&A...577A...3S} have analyzed 
the simulation of a group containing two galaxies like M31 and the Milky Way,
and while M33 is considerably less massive then the Milky Way, the simulation 
might still have some application to the M31-M33 region.  They find that the hot gaseous 
halos of the M31 and Milky Way analogs overlap and that they evolve to occupy the same filament, leading to 
an excess of neutral gas between them that forms around z $\sim$ 1 and 
persists to the present.  Clouds like those observed here might then condense in 
the filament. The current simulations 
do not have the resolution to detect anything as small as the M31-M33 clouds, and it is not clear that M31 and M33 
would have similar overlapping gas halos, but these results are encouraging that intra-group gas might be a natural 
feature of systems like the Local Group.

In view of the evidence that M31 has a massive CGM, as discussed in the next section, 
 it will be important to determine if the direction of M33 is enhanced in total material, 
as would be suggested by this scenario.

\subsection{Condensations in the M31 Circumgalactic Medium}

While the presence of the clouds between M31 and M33 is unexpected, the total mass involved is not 
large compared to the baryonic mass at that location in the halo of M31 as determined from recent 
measurements.   The 
observed \NHI\  averaged over our field is only $9 \times 10^{16}$ \cmm.
\citet{Lehner2015} have studied the extended CGM of M31 through measurement of UV absorption lines 
against background QSOs out to a projected distance $>500$ kpc, an area that covers not only the M31-M33 
clouds, but the galaxy M33 as well.   They find that the total gas mass in the M31 CGM may be  $>10^{10}$ 
M$_{\odot}$  with  an ionization fraction $>90\%$.   The Lehner et al.~radial column density profile of SiII evaluated at the 
projected distance $\rho \approx 100$ kpc of the M31-M33 field predicts N$_{\rm SiII} = 2.1 \times 10^{13}$ 
 \cmm, which implies an average total column density  ${\rm N_H} = 6 \times 10^{17}\ \rm{(Z_{\odot}/Z)}$~\cmm.
 Thus,  if the CGM of M31 has a sub-solar metallicity, even if it is  $>90\%$ ionized, 
it would have a neutral component of similar magnitude  to the average \NHI\ of the clouds over our field.

The tight kinematic pattern of the M31-M33 clouds suggests that they are not spread along the several 
hundred kpc path through the M31 halo, so arguments about their origin in the M31 CGM are suggestive 
at best. The \citet{Lehner2015} measurements, however, allow the possiblity that the
existence of clouds at $\rho = 100$ kpc from M31 does not require a major 
enhancement to the mass or density of 
its CGM, but could result from a restructuring or phase change in material already present, triggered, 
perhaps by the passage of a satellite as suggested by \citet{2013MNRAS.436.2096S}, or by 
a concentration of mass in a plane of satellites.  

The so called ``galactic fountain" model of gas accretion 
\citep[e.g.][]{1980ApJ...236..577B, 1976ApJ...205..762S} states that supernovae kick material above the disk of a spiral galaxy, 
which then rains back down onto the disk as it cools. 
More recent work has shown that this fountain material can cause the surrounding halo material to 
condense and fall onto the disk as well \citep{2008MNRAS.386..935F, 2015MNRAS.447L..70F}. 
While the M31-M33 clouds are too far away from either galaxy to be triggered by a galactic fountain,
 it is possible that hot gas could be triggered to condense by other external sources, such as
  the motion of a satellite galaxy through the CGM.  Such a satellite galaxy would presumably need to contain at least 
some cold gas to serve as a trigger.  It would be interesting to evaluate this mechanism in the 
context of the M31 CGM. 

\section{SUMMARY}
 
Prompted by the discovery of extended regions of very faint \HI that 
appear to form a partial bridge connecting M31 and M33 \citep{2004A&A...417..421B}, 
we have measured  21cm \HI emission in a 12 square-degree region between  M31 
and M33  with the Green Bank Telescope at $9\farcm1$ 
angular resolution reaching
$5\sigma$ limits on \NHI\ of $3.9 \times 10^{17}$ \cmm for a 30 \kms line.   
Sensitive observations were also made at 18 locations on a grid to the north of M31.

The new data confirm and extend 
the basic picture derived from the preliminary data presented in Paper I:
 in this region between M31 and M33  the \HI is 
largely if not entirely contained in  discrete neutral clouds 
that each have  M$_{\rm HI}$ reaching a few $10^5$ M$_{\Sun}$, 
lying at a projected distance $\approx 100$ kpc from M31.
 We do not find any evidence for 
a more diffuse component of \HI, and attribute our claim for this in Paper I to 
systematic instrumental baseline effects at the level of a few mK.  We measure 
only 63\% of the \HI mass found by BT04 in this region. While we present a possible explanation for this discrepancy (\S \ref{sec:neutralmass}), its origin is uncertain.
The clouds appear to be spatially and kinematically independent from each other and can 
have velocities that differ by $>100$ \kms over projected distances $\sim 10$ kpc. Our \ion{H}{1} mass limits of $\rm{\sim 10^4~M_\Sun}$ are lower, by an order of magnitude or more,
than other surveys of galaxy groups \citep[e.g][]{2006MNRAS.371.1617A, 2008AJ....135.1983C}. Thus these objects may represent a new, previously undetected population.

The clouds have a dynamical mass  nearly a thousand times their \HI mass and strong limits 
on any stellar component, making it unlikely that they are part of the dwarf galaxy 
system of the Local Group. Indeed, dwarf galaxies near large spirals in the Local Group completely lack detectable \HI 
\cite[][R. Beaton private communication]{Grcevich2009, 2014ApJ...795L...5S,Westmeier2015}. 
The clouds have kinematics more 
similar to the systematic velocity of the galaxies than to the HVC system of M31 and M33, 
but the clouds have \HI properties like those of the M31 HVCs and the class
of ultra-compact HVCs \citep{2005A&A...432..937W, 2008MNRAS.390.1691W, 2013ApJ...768...77A}.

Numerical simulations of the evolution of the Local Group produce regions of neutral gas
between the major galaxies that may be analogs to the detected clouds 
\citep{2014MNRAS.441.2593N} though considerably larger and more massive.  
If M31 has the very extensive circumgalactic medium (CGM)
recently proposed by \citet{Lehner2015}, then it contains $\sim10$ times the 
column density of gas needed for formation of the clouds at their projected radius.  The clouds might then 
be condensations in the $90\%$ ionized M31 CGM marking a past interaction with one or 
more of the dwarf galaxies \citep{2013MNRAS.436.2096S}.  
It will be critical in understanding the clouds to have an accurate census of the CGM of M31 
to determine if the M31-M33 direction is indeed a region of enhanced total mass.

Our results to the north-west of M31 are still too incomplete to determine if we have detected anything like 
the population of clouds that exist between M31 and M33, but the data do suggest that 
the  HVC population of M31 extends to $\rho \approx 100$ kpc in the north-west, much further than previously known. 
A similar extension to the south-east is not observed. 
The detection of \HI outside the BT04 contours at $\delta = +47\arcdeg$ would 
imply that we are seeing very small angular sized features that are beam-diluted in the BT04 measurements.  
Complete mapping of this area with the GBT is underway to resolve the discrepancy.

\acknowledgements
\vspace{-1.0em}
We thank Rachel Beaton for sharing results of her GBT survey of M31 dwarf galaxies before publication. S.A.W. acknowledges partial support from the student observing support grant
(GSSP11-012) provided by the NRAO.  D.J.P. and S.A.W. acknowledge partial support from NSF CAREER grant
AST-1149491.

\bibliography{Bibliography}{}

\end{document}